\documentclass[aps,pre,showpacs,amsmath,amssymb,amsfonts,lengthcheck,twocolumn,longbibliography,superscriptaddress]{revtex4}
 
\usepackage{changes} %to highlight revisions for peer review
\usepackage{ulem} % Paquete necesario para tachar texto
 \usepackage[utf8]{inputenc}
\usepackage{graphicx}
 \usepackage{subfigure}
\usepackage{amsthm}
\usepackage{mathtools}
\usepackage{amsmath}
\usepackage{verbatim}
\usepackage{dcolumn}% Align table columns on decimal point
\usepackage{bm}% bold math
\usepackage{color}
\usepackage[colorlinks=true,citecolor=blue,linkcolor=blue,urlcolor=blue]{hyperref}%
   \usepackage{xcolor}
\usepackage{physics}
\usepackage{dsfont}
 \usepackage{todonotes}
 
\usepackage{longtable}
 \usepackage{xargs}                      % Use more than one optional parameter in a new commands
\newcommand{\rescale}[2][1]{\resizebox{#1\textwidth}{!}{$#2$}}
\NewDocumentCommand\eq{ms}{%
  \IfBooleanTF{#2}{\begin{align}#1\end{align}}{\begin{align}
#1
\end{align}}
}
\newcolumntype{M}[1]{>{\centering\arraybackslash}m{#1}}
\definecolor{rojo}{RGB}{255, 0, 0}

\definecolor{bastian}{RGB}{136,0,123}

\begin{document}
\title{{Entropy, entanglement and susceptibility of three qubits near quantum criticality}}

\author{Bastian Castorene}
\email{bastian.castorene.c@mail.pucv.cl}
\affiliation{Instituto de Física, Pontificia Universidad Católica de Valparaíso, Casilla 4950, 2373223 Valparaíso,
Chile}
\affiliation{Departamento de Física, Universidad Técnica Federico Santa María, 2390123 Valparaíso, Chile}
\author{Francisco J. Peña} 
\affiliation{Departamento de Física, Universidad Técnica Federico Santa María, 2390123 Valparaíso, Chile}

\author{Ariel Norambuena} 
\affiliation{Departamento de Física, Universidad Técnica Federico Santa María, 2390123 Valparaíso, Chile}

\author{Sergio E. Ulloa} 
 \affiliation{Department of Physics and Astronomy and Nanoscale and Quantum Phenomena Institute, Ohio University, Athens, Ohio 45701, USA}
\author{Cristobal Araya} 

\affiliation{Departamento de Física, Universidad Técnica Federico Santa María, 2390123 Valparaíso, Chile}
 \author{Patricio Vargas}
\affiliation{Departamento de Física, Universidad Técnica Federico Santa María, 2390123 Valparaíso, Chile}

\date{\today}

\begin{abstract}
 
In this work, we investigate a system of three entangled qubits within the XXX model, subjected to an external magnetic field in the $z$-direction and incorporating an anisotropy term along the $y$-axis. We explore the system’s thermodynamics by calculating its magnetic susceptibility and analyzing how this quantity encodes information about entanglement. By deriving rigorous bounds for the susceptibility, we demonstrate that violations of these bounds serve as an entanglement witness. Our results show that anisotropy enhances entanglement, extending the temperature range over which it persists. Furthermore, when the system is traced over the degrees of freedom of two qubits, the reduced entropy of the remaining single qubit corresponds to that of a system in a thermal bath with temperature $T > 0$ K.

\end{abstract}

\maketitle

\section{Introduction}

The field of quantum technologies has undergone significant development, driven by the ability to create and control qubits using %various 
different experimental techniques.  {Among} the first prototypes were molecules where specific atoms served as effective two-level systems  {and similarly nuclear spins qubits were successful}. While {these} {approaches} laid the groundwork, {they} evolved significantly with the development of more precise and scalable techniques for multiqubit platforms~\cite{WOS:000324304400012,WOS:001156816000001,WOS:001027763900001,sergio1}.

Recent breakthroughs have expanded the range of materials and methodologies employed to fabricate qubits, including superconducting circuits~\cite{WOS:000952021200013, WOS:000492991700045}, trapped ions~\cite{WOS:000439545700001,WOS:000429529600002,WOS:000416520400037,WOS:000828441600002}, photons~\cite{WOS:000804867100011,  WOS:000599959400045} and semiconductor quantum dots~\cite{WOS:000744666700023, WOS:000744666700025, WOS:000744666700024}. Each of these platforms offers unique advantages in terms of coherence times, error rates, and scalability, contributing to the rapid progress in quantum computing, quantum communication, and quantum sensing \cite{WOS:001027763900001,WOS:000492991700045,WOS:000439545700001}.

Furthermore, the refinement of quantum error correction techniques and the development of quantum algorithms are pivotal in harnessing the full potential of these qubit systems. As research continues to address the challenges associated with qubit stability and manipulation, the prospects for practical quantum technologies become increasingly promising. Temperature plays a fundamental role in stability issues, affecting the entanglement properties that make qubit devices useful for their quantum properties \cite{WOS:000169822900048, WOS:000304905300040, WOS:000256441100041, WOS:000315709900039, WOS:000279014400020, WOS:000256528500006, WOS:000253764100063,VonNeumann1932}.
The study of a few interacting qubits has provided valuable insights into their thermal stability %concerning temperature 
and has enabled novel proposals for quantum engines driven by magnetic or electric fields \cite{WOS:001230903700002,WOS:001195261700001,WOS:000577236900006,xu2024universal,matos2023quantum}. In this context, this research focuses on a system of three interacting qubits using the Heisenberg XXX model in an external magnetic field along the $z$-axis, with magnetic anisotropy along the $y$-axis. By utilizing the 
%system's density matrix in 
the ground state ($T=0$ K), we investigated the effects of decoupling one qubit from the rest of the system by tracing over the states of the other two qubits. By examining the reduced entropy, we found that a single qubit exposed to the external magnetic field exhibits an effective temperature $T > 0$ K, even though the entire system remains in its ground state \cite{Castorene,technologies11060169,WOS:A1993LQ10700003,WOS:000317186000001,zhou2017quantum,mikitik2019magnetic,hurst2022non,bergholtz2021exceptional,Bell1, Werner1, Werner2}. 
 
{Our analysis delves into the magnetic behavior of the system, focusing on its ferromagnetic properties and the influence of anisotropy. Particular emphasis is placed on magnetic susceptibility}, exploring the temperature ranges within which the system exhibits entanglement. Additionally, we derive a simplified general expression for non-entangled states, using averages calculated from the tensor product of density matrices based on a susceptibility threshold at each temperature. This relationship aids in identifying the conditions under which the qubits remain either entangled or non-entangled, offering valuable insights into the thermal and magnetic characteristics of the system and their connection to entanglement \cite{WOS:000234347200003,WOS:000238204500001,WOS:000244532600066}.
{The ground-state properties of an interacting quantum system can abruptly change when a non-thermal control parameter, such as an external magnetic field, is tuned to a critical value in the limit of absolute zero temperature \cite{qpt1,qpt2,qpt3,qpt4,qpt5,qpt6}. This phenomenon is referred to as a quantum phase transition (QPT).

In the literature, significant efforts have been dedicated to classifying and analyzing these transitions, often focusing on the role of ground-state level crossings and singularities in thermodynamic or correlation functions \cite{qpt1,qpt7,qpt8,qpt9}. In this work, we adopt the Ehrenfest classification scheme, which identifies quantum critical points (QCPs) through the non-analytic behavior in the n-th derivative of two-point spin correlations at the critical parameter in the zero-temperature limit \cite{qpt10}. In particular, a discontinuity in the first derivative with respect to the magnetic field evidences a first-order QPT, highlighting the sensitive interplay between quantum fluctuations and criticality.
}

\section{Model}
The working substance consists of three qubits described by a Heisenberg XXX model with nearest-neighbor interactions, an external magnetic field along the \(z\) direction, and magnetic anisotropy term along the \(y\) direction.

To describe the system, we use the Pauli matrix operator \(\boldsymbol{S}_i = (1/2)(\hat{\sigma}_i^x, \hat{\sigma}_i^y, \hat{\sigma}_i^z)\), representing the spin at the \(i\)-th qubit in $\hbar$ units. The
total \(y\)-spin operator on the three qubits is \(S_y = (1/2)(\hat{\sigma}_1^y + \hat{\sigma}_2^y + \hat{\sigma}_3^y)\).
The nearest neighbor exchange coupling constant \(J\) will be used as a unit of energy and is set to \(-1\) for a ferromagnetic configuration and \(1\) for an antiferromagnetic configuration. The variable \(\alpha\) is binary, with {zero value} for chain topology and one for the ring topology connecting qubit one with qubit three. The external magnetic field \(B\) points along the \(z\)-axis, and \(K\) represents the anisotropy term. When the anisotropy is positive (\(K > 0\)), there is easy-plane anisotropy, which maximizes magnetization in the \(x\)-\(z\) plane. When the anisotropy is negative (\(K < 0\)), there is easy-axis anisotropy, which maximizes magnetization along the \(y\)-axis. Therefore, the Hamiltonian describing the system is given by
\begin{equation}\label{H}
    \mathcal{H} = {J}\qty(\sum_{i=1}^2 \boldsymbol{S}_i \cdot \boldsymbol{S}_{i+1}+ \alpha \qty(\boldsymbol{S}_1\cdot \boldsymbol{S}_3))+ K \boldsymbol{S}_y^2 + B\sum_{i=1}^3  {S}_i^z
\end{equation}
Note that $J,K$ and $B$ have the same energy units. 
The energies $E_n$ of the system are:
\eq{
\begin{aligned}
& E_1=\frac{1}{4}(-2 B-3 \alpha J+K), && E_2=\frac{1}{4}(2 B-3 \alpha J+K), \\
& E_3=E_1+J(\alpha-1),&& E_4=E_2+J(\alpha-1), \\
& E_5=\frac{1}{4}\left(2 B+J_{\mathrm{eff}}-4 P_{-}\right) && E_6=\frac{1}{4}\left(2 B+J_{\mathrm{eff}}+4 P_{-}\right) \\ &E_7=\frac{1}{4}\left(J_{\mathrm{eff}}-2 B-4 P_{+}\right), && E_8=\frac{1}{4} \left(J_{\mathrm{eff}}-2 B+4 P_{+}\right),
\end{aligned}\label{eigenenergies}
}
 \noindent where the variables used are \(J_{\text{eff}} = (2+\alpha)J + 5K\) and \(P_\pm = \sqrt{B^2 \pm  BK +  K^2}\). The other auxiliary variables related to the eigenvectors are detailed in Appendix~\ref{variable_change}.
 In the usual \(\sigma_z\) basis: \(\ket{0}=[1,0]^T\), \(\ket{1}=[0,1]^T\), {the} corresponding eigenvectors in the three-qubit computational basis are given by:

\eq{{\begin{aligned}
\ket{\psi_1} & =   \frac{1}{\sqrt{2}}\qty(  \ket{110}-\ket{011})\\
\ket{\psi_2} & =   \frac{1}{\sqrt{2}}\qty(\ket{100} - \ket{001})\\
\ket{\psi_3} & =   \frac{1}{\sqrt{6}} \qty(\ket{110}+\ket{011}-2 \ket{101})\\
\ket{\psi_4} & =  \frac{1}{\sqrt{6}} \qty(\ket{100}+\ket{001}-2 \ket{010}) \\
\ket{\psi_5}&= L_0^-\qty( \frac{4R_0^-}{ K} \ket{000} + \ket{011}+ \ket{101}+\ket{110} )\\
 {\ket{\psi_6}} &  {= \frac{L_0^{-}}{\sqrt{3}}\left(\frac{4R_0^{-}}{K}(|011\rangle+|101\rangle+|110\rangle)-3|000\rangle\right)} \\
\ket{\psi_7} &=  \frac{L_1^+}{\sqrt{3}} \qty(3\ket{111} - \frac{ 4R_1^+}{K} \qty(\ket{100}+\ket{010}+\ket{001}))     \\
\ket{\psi_8}&= \frac{L_0^+}{\sqrt{3}} \qty(3\ket{111} - \frac{4 R_0^+}{K} \qty(\ket{100}+\ket{010}+\ket{001}))    
\end{aligned}}\label{eigenvectors}}
{Using the standard $\sigma_z$  basis representation, $|0\rangle=[1,0]^T$ and $|1\rangle=[0,1]^T$.}
{An immediate consequence of the system's topology on the energy spectrum is that, in the ring arrangement (\( \alpha = 1 \)), degeneracies such as \( E_1 = E_3 \) and \( E_2 = E_4 \) always occur. This situation contrasts markedly with the chain arrangement (\( \alpha = 0 \)), where no such degeneracies are present for \( B > 0 \). Moreover, when \( B = 0 \), there exists a fourfold degeneracy involving the energies \( E_1 \) and \( E_2 \). In this scenario, the states \( \ket{\psi_1} \) and \( \ket{\psi_2} \) form Kramers pairs.}
The exchange coupling constant {determines the nature of system ground states}, resulting in a ferromagnetic ground state (\(J = -1\)) with {maximal spin for positive magnetic field described by $\ket{\psi_7}$} In contrast, the antiferromagnetic coupling {$(J=1)$ produces a ground state with} double degeneracy for the same magnetic field values.

\begin{figure*}
\centering
\includegraphics[width=.90\textwidth]{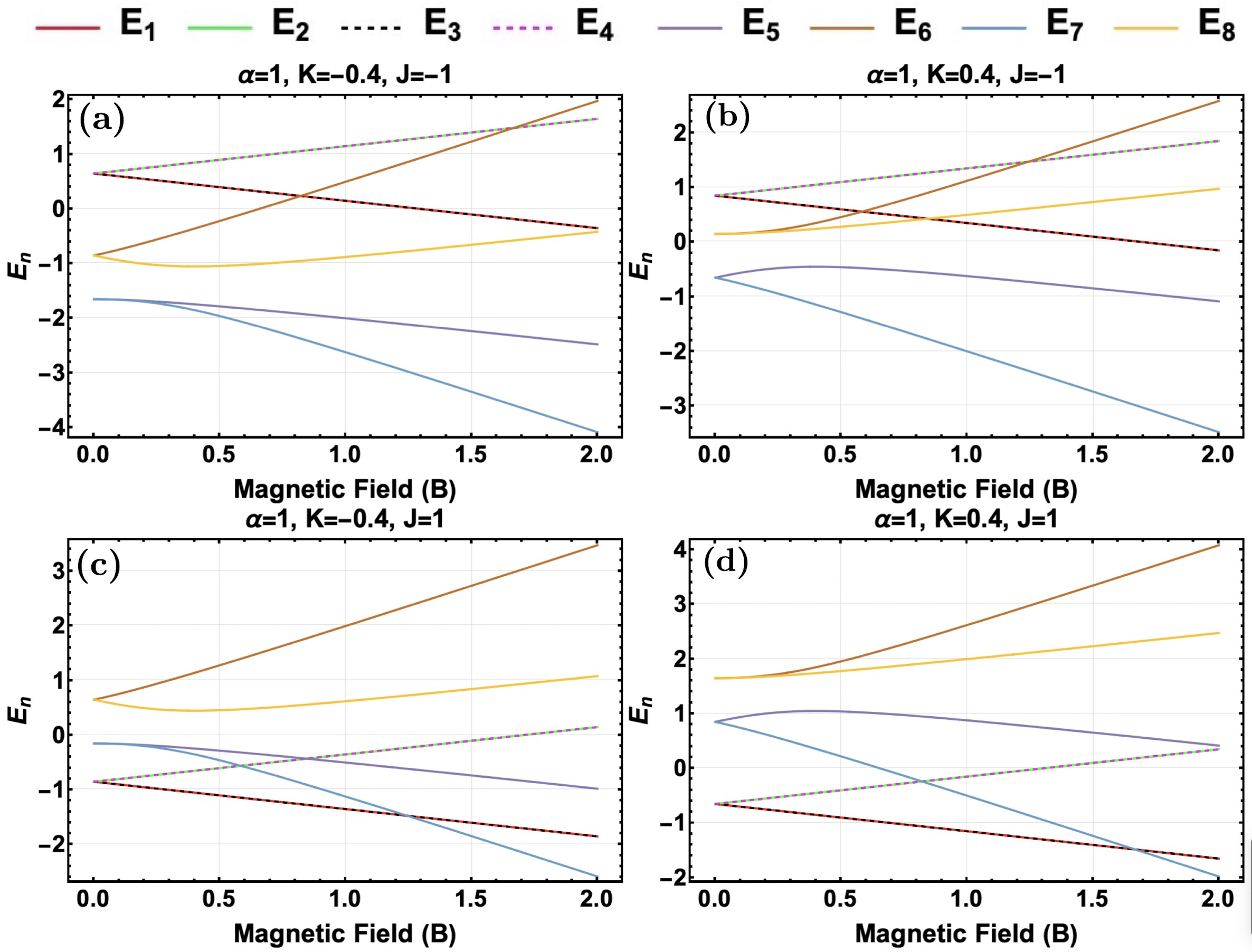}
\caption{The energy spectrum of the ferromagnetic $(J=-1)$ and antiferromagnetic $(J=1)$ exchange for a ring system $\alpha=1$, across different anisotropies $K$ as a function of the magnetic field $B$. In (a) and (b), for the ferromagnetic system $(J=-1)$, it can be observed that the ground state {$E_7$} never intersects any other energy level when $B > 0$. In (c) and (d), for the antiferromagnetic system $(J=1)$, a {level} crossing occurs at the energies $E_1 = E_3 = E_7$. The values of anisotropy shift this crossing to the right for positive anisotropy and to the left for negative anisotropy. Degeneracies $E_1 = E_3$ and $E_2 = E_4$ {are always present} for the ring arrangement $(\alpha=1)$
 }
\label{energias_ferro}
\end{figure*}

In Figure~\ref{energias_ferro} (a) and (b), for the ferromagnetic {coupling} (\(J=-1\)), it can be observed that the ground state never crosses any other energy level when \(B > 0\).

Figure~\ref{energias_ferro} (c) and (d) show that the ground state at \(B=0\) is {fourfold}-degenerate and represented by a linear superposition of states \(\left|\psi_i\right\rangle\), with \(i \in \{1,2,3,4\}\). With a slight increase in the magnetic field \((0 < B < B_{\text{crit}})\), \(B_{\text{crit}} = (-K +\sqrt{9 J^2 + 12 J K + K^2})/2\), the ground state splits into the {entangled} states \(\left|\psi_1\right\rangle\) and {\(\left|\psi_3\right\rangle\)}, forming a doubly degenerate ground state. In contrast, for \(B > B_{\text{crit}}\), the {non-entangled maximal spin triplet state} \(\ket{111} = \left|\psi_7\right\rangle\) becomes the ground state. This {illustrates} that the ground state properties can be manipulated through external magnetic field adjustments and anisotropy values.
To characterize the system thermodynamics, we use the density operator in thermal equilibrium defined by
\begin{equation}
    \rho = \frac{e^{-\beta \mathcal{H}}}{\Tr{e^{-\beta \mathcal{H}}}} = \sum_{i=1}^8 P_i \ket{\psi_i}\bra{\psi_i}
    \label{rho_total}
\end{equation}
where $P_i= e^{\left(-\beta {E_i}\right)}/Z$ corresponds to the Boltzmann occupation probabilities of the eigenstates $\ket{\psi_i}$ with $Z = \sum_{i=1}^8 \exp{ -\beta {E_i}}$ being the canonical partition function of the system. Here, we use the Boltzmann constant $k_B = 1$ as a convention for our calculations and therefore $\beta=\frac{1}{T}$. This matrix operator given by Equation \eqref{rho_total} describes the eight-level system of three interacting qubits and satisfies the conditions $\rho = \rho^\dagger$, $\Tr{\rho} = 1$, and $\sum_i \bra{\phi_i}\rho \ket{\phi_i} \geq 0$. 
{In the next section, we will discuss the role of spin correlations in the thermodynamical context to characterize the quantum phase transition and the nature of the critical magnetic field $B_{\text{crit}}$.
\subsection{Spin-spin correlations and Quantum phase transition}
 The QPT is expected to manifest in spin correlations as the ring’s degenerate ground state, represented by the states $\left|\psi_1\right\rangle$ and $\left|\psi_3\right\rangle$, as $B \rightarrow 0$. As $B$ increases beyond the critical magnetic field value $B_{\text{crit}}$, the degeneracy is broken, and the system transitions into a fully aligned ferromagnetic state, $\left|\psi_7\right\rangle$, characterized by maximal spin alignment at low temperatures.
\begin{figure*}
    \centering
    \includegraphics[width=.85\linewidth]{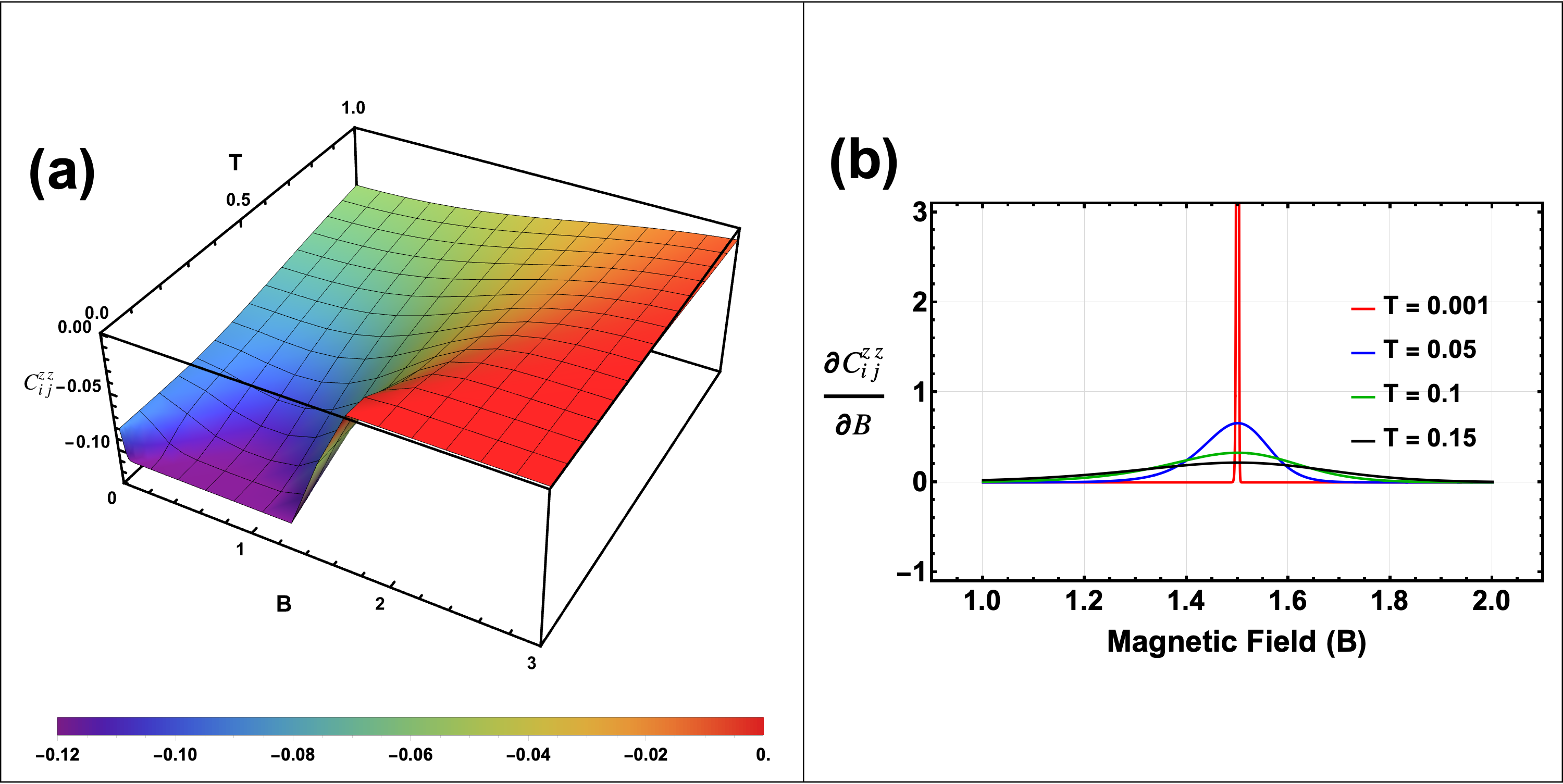}
    \caption{{Variation of the \(zz\) spin-pair correlations in the antiferromagnetic ring system (\(J = \alpha = 1\)) at \(K = 0\) under thermal equilibrium. Panel (a) depict the correlations as functions of temperature (\(T\)) and magnetic field (\(B\)). Panel (b) illustrate the first-order derivatives of the \(zz\) correlations at selected temperatures: \(T = 0.001\) (red solid line), \(T = 0.05\) (blue solid line), \(T = 0.1\) (green solid line), and \(T = 0.15\) (black solid line).  The quantum phase transitions at zero temperature are located in the  maximum peaks at the quantum critical points (QCPs) at the fixed magnetic field $B_{\text{crit}}$.
       }}
    \label{CorrelacionesYDerivadas}
\end{figure*}
To observe this transition, consistent with the definition of a quantum phase transition, we introduce the two-point spin correlation function:
\begin{equation}
C_{ij}^{kl} = \expval{ S_i^k S_j^l } - \expval{S_i^k} \expval{S_j^l}, \label{spin_correlation_function}
\end{equation}
where the expectation values are computed using the density matrix at thermal equilibrium \eqref{rho_total} as $
\expval{\mathcal{O}} = \Tr{\rho \, \mathcal{O}}$.
Here, the qubit site labels and the spin components are $\qty{i, j} \in \qty{1, 2, 3}$, $\qty{k, l} \in \qty{x, y, z}$.

%Let us assume that the pair-wise qubit state is described by the pure (uncorrelated) state in \eqref{two_qubit_rho}. In this case, the correlation function evaluates to $C_{ij}^{kl} = 0$, indicating the absence of correlations along any degree of freedom. Conversely, any nonzero value of $C_{ij}^{kl} \neq 0$ implies the presence of spin correlations within the system.

For the ring system $(\alpha = 1)$, the two-qubit pairs are indistinguishable particles, and each particle interacts symmetrically with its nearest neighbors. Therefore, the correlation function satisfies $C_{ij}^{kl} = C_{ji}^{kl}$, forming a symmetric tensor under the exchange of indices $\qty{i, j}$. This symmetry reduces the total number of independent calculations from 27 to only the upper diagonal terms, leaving 9 elements. Furthermore, in the specific case where the anisotropy $K = 0$, the system’s symmetry ensures that correlations with $k \neq l$ satisfy $C_{ij}^{kl} = 0$ \cite{Castorene, qpt9}, thereby further reducing the number of independent elements to only three.

The presence of the exchange constant $(J)$ and anisotropy $(K)$ within the system acts as a source of entanglement, inducing specific degrees of spin correlation between the qubits.

}
{In Fig.~\ref{CorrelacionesYDerivadas}~(a), the \(zz\)- and \(xx\)-correlations of an antiferromagnetic ring system \((J = \alpha = 1)\) with anisotropy \(K = 0\) are plotted as functions of the magnetic field \(B\) and temperature \(T\). As \(T \to 0\), both correlations exhibit a pronounced change near \(B_{\text{crit}} = 3/2\). This behavior arises from the crossing of ground-state energies, leading to a discontinuity in the first derivative of the ground-state energy with respect to \(B\), as shown in Fig.~\ref{energias_ferro}~(b). Near zero magnetic field, a symmetry breaking occurs, transitioning the system from a tetradegenerate ground state to a doubly degenerate one. This transition is reflected in the \(zz\)-correlation, which decreases from a higher to a lower value.

In the \(zz\)-correlation, when \(B \neq 0\), the system transitions from a doubly degenerate one, with a reduction in spin correlation, and eventually to a single separable state characterized by maximal spin alignment. These symmetry changes are indicative of first-order quantum phase transitions \cite{qpt1,qpt2,qpt4}. 

Notably the $zz$ correlation spotlight the critical point associated with the quantum phase transitions for our model even at finite T, because the derivative with respect to the field has its maximum at the same magnetic field $B=B_{crit}=3/2$ for distinct temperatures. Therefore, for our system, we can see that $zz$ spin-spin correlation  detects the QPT at finite T, at $B >0$.

Moreover, since the spin-spin correlation functions are directly linked to thermodynamic quantities such as magnetization, specific heat, and entropy, the $zz$ correlation can be accessed through measurements performed at finite temperatures \cite{technologies11060169}. This has important implications for the experimental characterization of quantum phase transitions (QPTs), especially in cases where temperatures low enough to directly observe a QPT cannot be achieved.

When anisotropy is present (\(K \neq 0\)), the critical field \(B_{\text{crit}}\) shifts to the right (left) for \(K > 0\) (\(K < 0\)) at zero temperature, following \(B_{\text{crit}} = {(-K + \sqrt{9 J^2 + 12 J K + K^2})}/{2}\), as shown in Fig.~\ref{energias_ferro}(c)-(d). While the position of the critical point changes, the nature of the quantum phase transition remains unaltered, retaining its classification as a first-order QPT.
}

{\section{One qubit effective model}}
{\subsection{Effective Hamiltonian}
    {We can construct an effective model by isolating a single qubit of the system that interacts solely with the external magnetic field in the \(z\)-direction at temperature $T$, due to the presence of an effective thermal bath and without any other interactions, such as exchange and magnetic anisotropic, obtaining:}
 \eq{\mathcal{H}_{\text{eff}}&= B S_z \qquad (+\text{Heat Bath})
 }
 The associated eigenenergies are:
 \eq{E_\text{eff}^\pm &= \pm \frac{B}{2} 
 }
 {The corresponding partition function and Helmholtz free energy {are} given by the usual canonical ensemble formulas}:
 \eq{Z_\text{eff}&= 2 \cosh\qty( \frac{B}{2T}) \implies F_\text{eff}= -T \ln \left[2 \cosh \qty(\frac{B}{2T})\right]
 }
 {The associated entropy of the isolated qubit is:
 \eq{S_\text{eff}=  \ln \qty[ 2\cosh \qty(\frac{B}{2T})]-\frac{B}{2T} \tanh \left(\frac{B}{2T}\right) \label{S_efff}}} 
 }
\subsection{Reduced ferromagnetic entropy}
{The definition of entropy that we will use throughout the work will be the one given by von Neumann \cite{VonNeumann1932}, which is provided by}
{
\begin{equation}
    S = -\Tr{\rho \ln \rho}.
    \label{generalentropy}
\end{equation}
}
{We will refer to \textit{the ferromagnetic von Neumann entropy} ($S_{g}$) to that one associated only with the ground state of the system and applicable for both geometries \(\alpha = \{0,1\}\). This entropy can be calculated by constructing the density matrix of the ferromagnetic ground state  \((\ket{\psi_7})\), or by considering the limit at zero temperature and $J<0$ using the density matrix of Equation \eqref{rho_total}. Accordingly, we have }
\[
\rho_g = \rho (T\rightarrow 0) =   \ket{\psi_7}\bra{\psi_7}.
\]
{Subsequently, we perform a partial trace over two qubits to reduce the system to a single qubit. Remarkably, the result of the partial trace is the same across all qubits, regardless of the tracing sequence, yielding \( \rho_g^{i|jk} \) for \( \{i, j, k\} \in \{1, 2, 3\} \) with \( i \ne j \ne k \), where \( \rho_g^{i|jk} = \Tr_j\left( \Tr_k\left( \rho_g \right) \right) \).}
{
\eq{\rho_g^{i|jk}= \mqty(\frac{2B+K+4P_+}{6 P_+} &0 \\ 0 & -\frac{2B+K- 2P_+}{6 P_+})\label{resultsreduce}
}
Following this, we calculate the von Neumann entropy obtaining:}
\eq{ S_g (K,B) &= - \Tr{ \rho_g^{i|jk} \ln\qty({\rho_g^{i|jk}})} \nonumber \\
& = \ln \qty(\frac{6P_+}{2B+K+4P_+}). \label{SGround}
}
This expression depends on the magnetic field \( B \) and the magnetic anisotropy \( K \), but it does not depend on the geometry parameter \( \alpha \) or explicitly on the coupling constant \( J \). The influence of \( J \) arises solely from the sign chosen to ensure that the ground state is ferromagnetic. If we retain \( J \) throughout the derivation, it simplifies and does not appear explicitly in the entropy expression.
   
\subsection{Effective Temperature}
 The isolated qubit temperature can be computed using {Equation \eqref{S_efff}}. This is achieved by identifying the point where the entropy function {of Equation \eqref{S_efff}} aligns with the groundstate effective entropy {of Equation} \eqref{SGround}, {which means \(S_{\text{eff}}(T,B) = S_{g}(K,B)\).} {This defines} the temperature corresponding to {the} isolated {qubit}. We obtain the following coupled equation for $T$ {in terms of $B$ and $K$:}
\eq{ \ln \qty[  2\cosh \qty(\frac{B}{2T})]-\frac{B}{2T} \tanh \left(\frac{B}{2T}\right) = \ln \qty(\frac{6P_+}{2B+K+4P_+})\label{Teffective}}
Eq.~\eqref{Teffective} is a transcendental equation where numerical calculations must be applied to find the temperature $T$.  
\begin{figure}
\centering
\includegraphics[width=.435\textwidth]{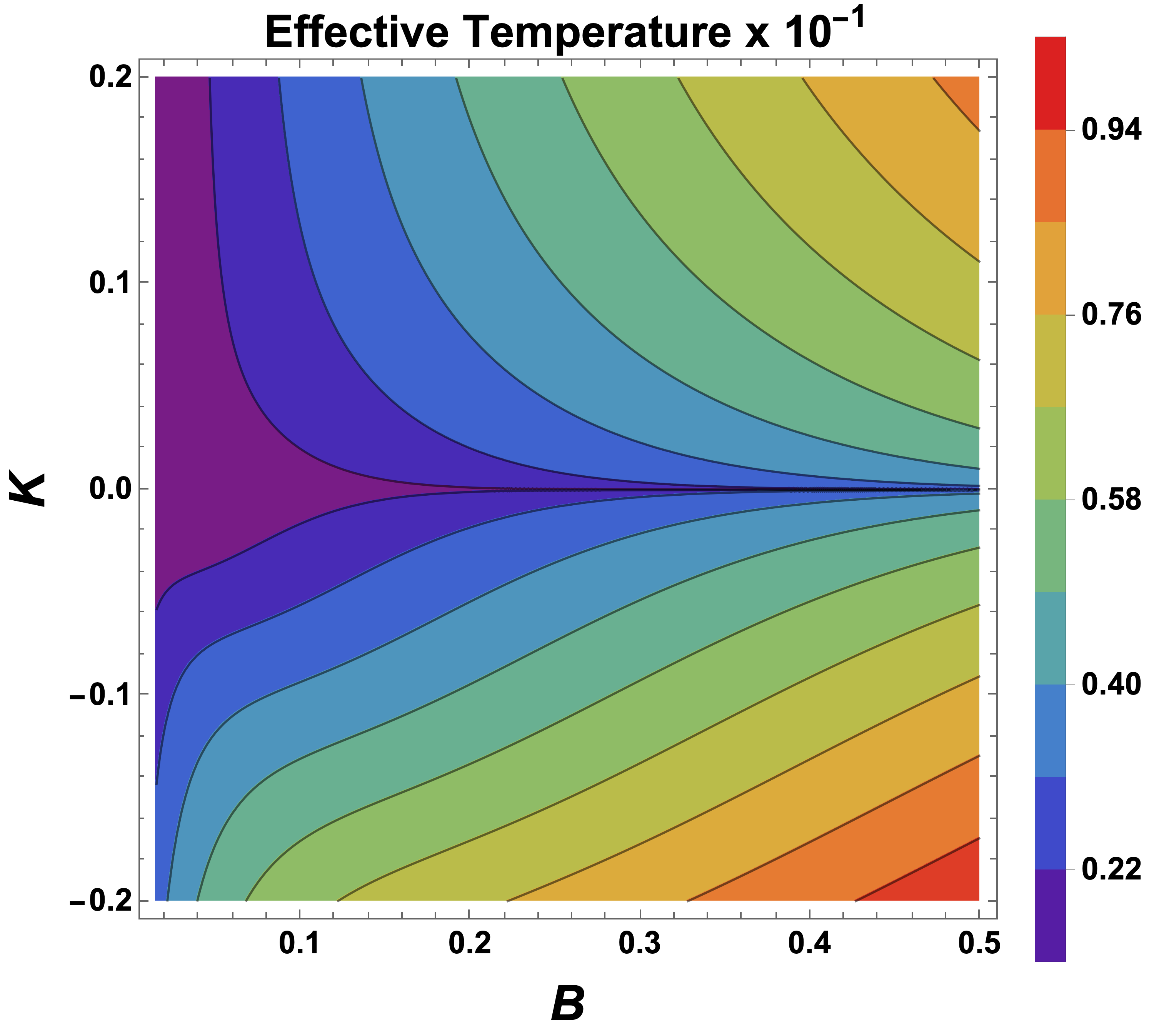}
\caption{Temperature map against Magnetic Field $B$ and Anisotropy $K$ (in units of energy), solved numerically using Eq.~\eqref{Teffective}. One can notice the different behavior for positive and negative values of $K$. {One can see that temperature increases as function of $B$ for $K\not =0$ and for $K=0$ we have zero temperature.} { Note that the magnetic field field is strictly $B>0$.}}
\label{TemperaturasEfectivas}
\end{figure}

In Figure \ref{TemperaturasEfectivas}, the behavior of the system under positive and negative anisotropy is shown, revealing an asymmetry with respect to the horizontal line at \(K = 0\). This arises from the distinct nature of anisotropy: \(K < 0\) corresponds to easy-axis anisotropy, while \(K > 0\) represents easy-plane anisotropy. 

 {For \( K = 0 \), note that the temperature is \( T  = 0 \). More generally, for any temperature, we can associate a functional relationship between \( K \) and the magnetic field \( B \). {Note that the magnetic field is strictly positive (\(B > 0\)). Consequently, the level \(\ket{\psi_6}\), which is degenerate with \(\ket{\psi_7}\) at zero field, will not participate. Since the entire system is at absolute zero temperature, it cannot be excited to this level.
}

In general, it is observed that the temperature increases as a function of \( B \) when \( K \neq 0 \). For negative values of \( K \) and a fixed temperature, \( K \) increases as \( B \) grows. In contrast, for positive values of \( K \), anisotropy decreases as the magnetic field \( B \) increases in order to maintain a constant temperature. 

In the case of the antiferromagnetic ring system (\( J >0 \)), the degenerate ground-state eigenvectors \( \ket{\psi_1} \) and \( \ket{\psi_3} \) are independent of the external parameters \( B \) and \( K \)
{To ensure this assertion holds, the magnetic field must be strictly positive (\(B > 0\)). Consequently, the levels \(\ket{\psi_2}\) and \(\ket{\psi_4}\), which are degenerate with \(\ket{\psi_1}\) and \(\ket{\psi_3}\) in the absence of a magnetic field, will not contribute. Furthermore, since the system is maintained at absolute zero temperature, transitions to these states are prohibited.
}However, we can prepare the states by using the parameter \( a^2 \), representing the probability of the state \( \ket{\psi_1} \). Therefore, the antiferromagnetic ground-state density matrix \( \rho_{\text{AF}} \) is given by:
\begin{equation}
\rho_{\text{AF}} = a^2 \ket{\psi_1}\bra{\psi_1} + (1 - a^2) \ket{\psi_3}\bra{\psi_3}. \label{rho_antiferro}
\end{equation}

From Eq.~\eqref{rho_antiferro}, the reduced density matrix for an isolated qubit is:
\begin{equation}
\rho_{\text{AF}}^{i|jk} = \begin{pmatrix}
c_{11} & 0 \\
0 & c_{22}
\end{pmatrix}, \label{rho_AF_reduced}
\end{equation}

where \( c_{11} = \frac{1 + 2a^2}{6} \) and \( c_{22} = \frac{5 - 2a^2}{6} \). The corresponding von Neumann entropy is then:
\begin{equation}
S_{\text{AF}}(a) = - c_{11} \log c_{11} - c_{22} \log c_{22}. \label{SAF}
\end{equation}

\begin{figure}
\centering
\includegraphics[width=.435\textwidth]{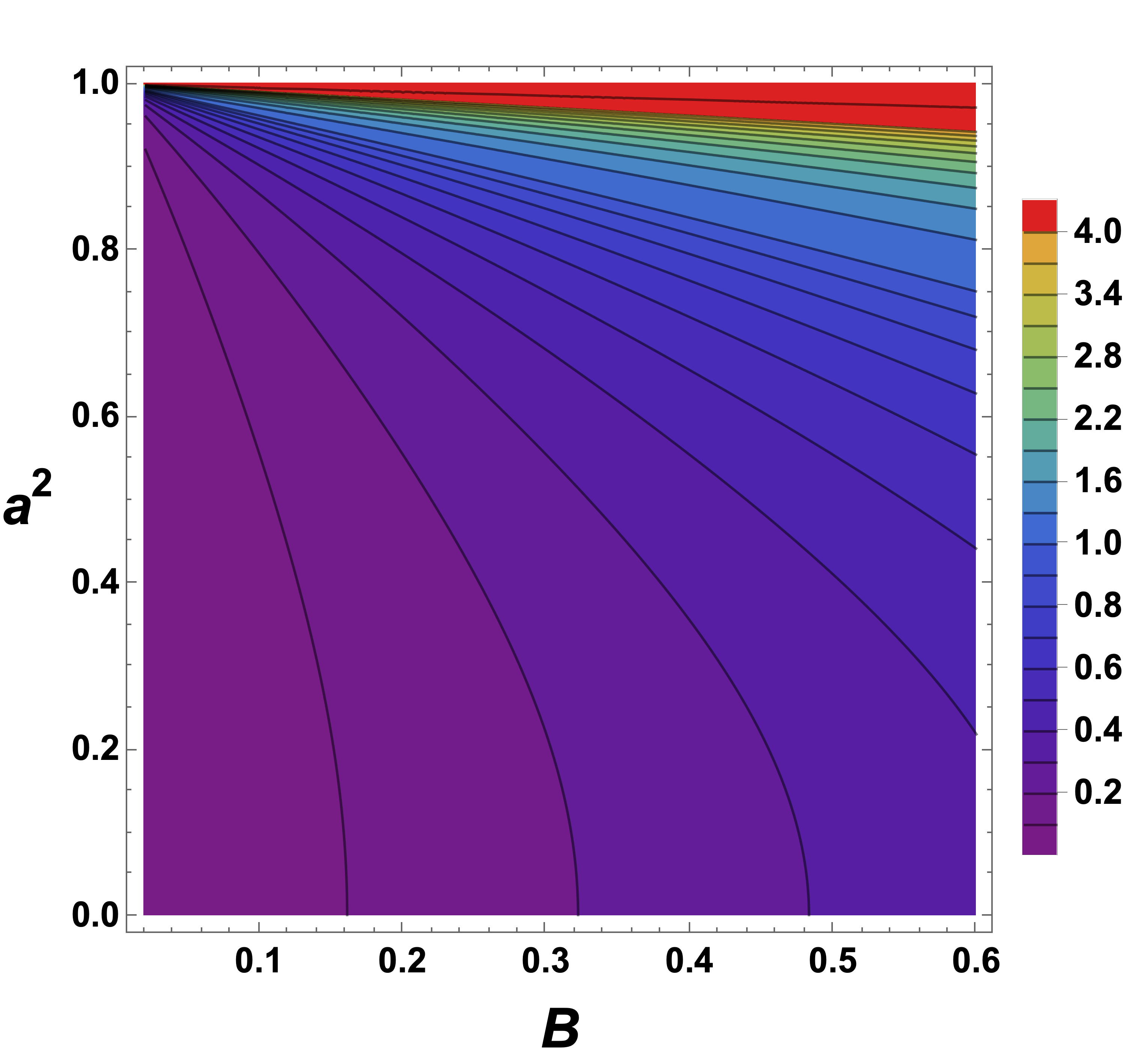}
\caption{{Effective temperature map against $B$  and $a^2$, It can be notice that temperature increases as function of $B$. As $a^2$ goes to 1, the temperature goes to infinity for every magnetic field greater than zero.}{The magnetic field is strictly $B>0$.}}
\label{AlphaBTemperature}
\end{figure}
Now, for a given magnetic field \( B \), we obtain the temperature for the isolated qubit by using Eqs.~\eqref{S_efff} and \eqref{SAF}. Figure~\ref{AlphaBTemperature} depicts the temperature map of the isolated qubit as a function of both the prepared ground-state parameter \( a^2 \) and the magnetic field \( B \). We observe that the temperature rapidly diverges as \( a^2 \) approaches to one and is independent of the magnetic field. Additionally, for a fixed \( a^2 \), increasing the magnetic field \( B \) results in a higher temperature. To achieve similar temperatures across varying magnetic fields, the value of \( a^2 \) must decrease accordingly. Notably, when \( a^2 = 0 \), the temperature is non-zero and increases with the magnetic field.}
\section{Magnetic Properties}
\subsection{Magnetization}
The magnetization of the
system can be analytically calculated using the canonical ensemble formulas and standard thermodynamic properties:
\eq{ M(J,\alpha,K,B,T)=-\left(\frac{\partial F}{\partial B}\right)_T=T \left(\frac{\partial \ln Z}{\partial B}\right)_{T} \label{Magnetizationeq}}
\begin{figure*}
\centering
\includegraphics[width=1\textwidth]{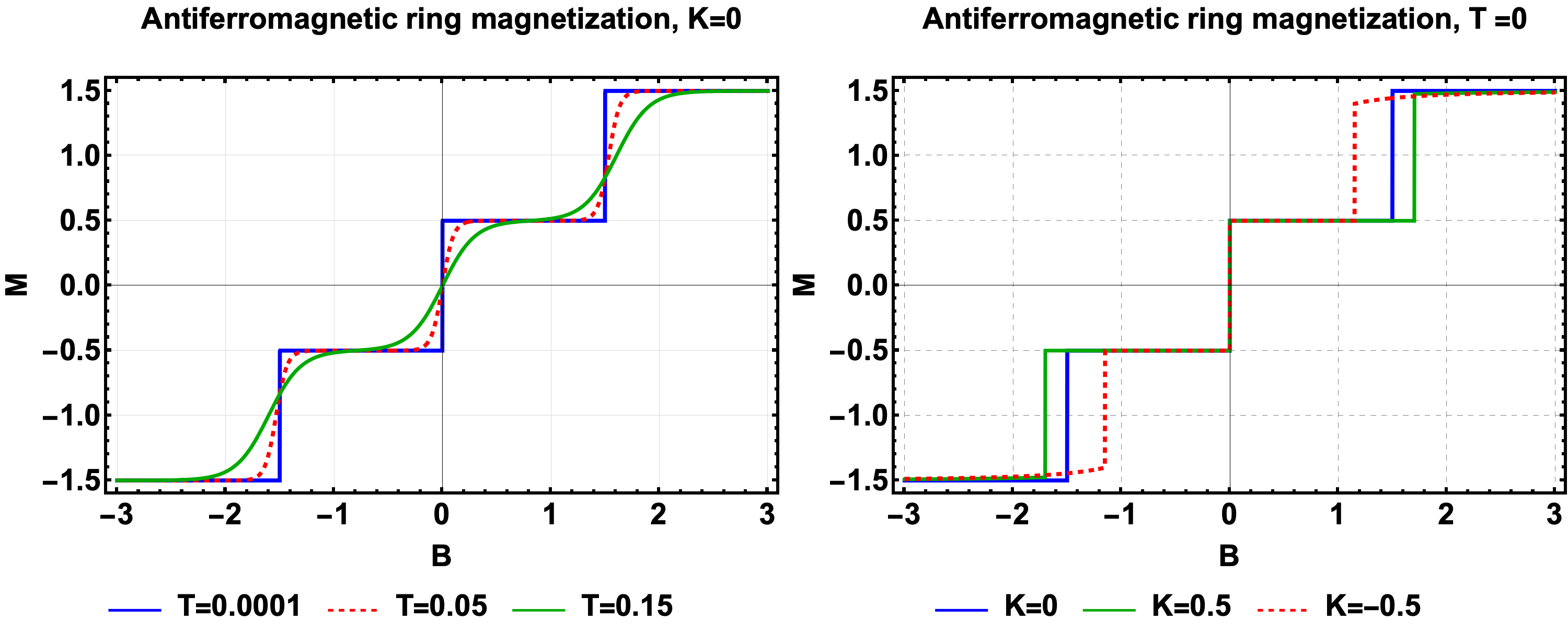}
\caption{ Antiferromagnetic $(J=1)$ {ring} ($\alpha=1$) magnetization against magnetic field (B). \textbf{(Left)} Magnetization as a function of the magnetic field at different temperatures with fixed anisotropy $K=0$. At zero temperature, the magnetization exhibits a ladder-like behavior, with steps occurring exactly at the quantum critical points {(QCPs) $B=0$ and $B=3/2$}. As the temperature increases, these steps become {noticeably} smoother. \textbf{(Right)} Magnetization as a function of the magnetic field at different anisotropies at fixed temperature $T=0$. The effects of anisotropy shift the {QCPs} to the right for positive values and to the left for negative values of $K$, causing the steps to occur earlier for negative anisotropy and later for positive anisotropy. }
\label{Magnetizaciones}
\end{figure*}
In Figure ~\ref{Magnetizaciones} (Left), {for the ring $(\alpha=1)$} one can see that at almost zero temperature, the antiferromagnetic magnetization with \(K=0\) exhibits a ladder-like behavior, starting at $B=0$ the first step has a width of {\(1/2\)}. This occurs because the system is frustrated: two spins are antiparallel to each other, while the third spin does not have a fixed behavior and is the first to align with the magnetic field. Precisely at the quantum critical point (QCP), the other two spins align in the same direction, creating a step of $1$. The effect of temperature smooths this behavior, resulting in a more classical appearance.
               
In Figure~\ref{Magnetizaciones} (Right), the magnetization behavior at zero temperature shows that the {anisotropy shifts} the jumps of the two antiparallel spins to the left if the anisotropy is negative or to the right if it is positive. Negative anisotropy results in a smoother transition near the QCP, giving the appearance that the system is at a higher temperature than it truly is in that region.
\subsection{Magnetic Susceptibility}
 {
 Magnetic susceptibility is a classical thermodynamic quantity that can be experimentally measured and serves as an entanglement witness. It can be derived from the magnetization expression Eq.~\eqref{Magnetizationeq} as follows:
\eq{\chi_z = T \eval{\left(\frac{\partial^2{\ln Z}}{\partial B^2}\right)_{T}}_{B=0} = \eval{\left(\pdv{M}{B}\right)_{T}}_{B=0}}
In the special case of a zero magnetic field, the magnetic susceptibility takes the following form:
\eq{\chi_z(J,\alpha,K,T)= \frac{1}{2T}\qty(\frac{1}{2}-\frac{1}{2K}\frac{K+3T-3e^{\frac{2K}{T}}(K+T)}{1+e^{\frac{2K}{T}}\qty(1+e^{\frac{3J}{2T}}+e^{\frac{J(1+2\alpha)}{2T}})})
\label{chiB0}
}
We analyze the high-temperature inverse susceptibility behavior of the system in this scenario, resulting in the form of the Curie-Weiss law:
\eq{1/\chi_z \xrightarrow{T \gg 1} \frac{4T}{3} + \frac{2J}{9}(2 + \alpha) \label{x_curie_high_T}}
We observe from the last Equation.~\eqref{x_curie_high_T}, that the susceptibility is independent of the anisotropy and depends only on the temperature, exchange constant $J$ and topology term $\alpha$. 

The inverse susceptibility against temperature for different exchange interactions is depicted in Figure~\ref{susceptability_temperature}, considering zero magnetic field and zero anisotropy. The antiferromagnetic case ${(J=1)}$ is shown as the red line, having the highest inverse susceptibility values, while the ferromagnetic case ${(J=-1)}$, shown as the blue line, has the lowest. 
{The paramagnetic scenario (\(J \to 0\)), represented by the black line, lies between the antiferromagnetic and ferromagnetic cases. At high temperatures (\(k_B T \gg \abs{J}\)), the antiferromagnetic extrapolation intersects the horizontal axis at a negative value, indicating a Néel temperature. The ferromagnetic extrapolation intersects at a positive value, corresponding to a Curie temperature, while the paramagnetic scenario intersects at the origin, marking a Langevin temperature. These extrapolated lines emphasize that all three systems converge towards a paramagnetic regime at sufficiently high temperatures, as thermal energy dominates over interaction strength.}

{In Appendix \ref{Entanglement_Inequality}, we show that the magnetic susceptibility as a function of the temperature of a non-entanglement system, has an upper bound given by
\eq{
\chi_z < \frac{1}{2T} \label{Nqubitineeq1}.
}
Therefore, the criterion for entanglement is met when the inequality in Eq.~\ref{Nqubitineeq1} is violated.
The green dashed line in Figure~\ref{susceptability_temperature}  represents the threshold given by Eq.~\eqref{Nqubitineeq1}. The curves below this line are non-entangled system, while those above are. Only the antiferromagnetic case lies above this line at $T \leq 3/(2\ln 3 )$, {as one would expect because of the entanglement.}}}
  
\begin{figure}
    \centering
    \includegraphics[width=1\linewidth]{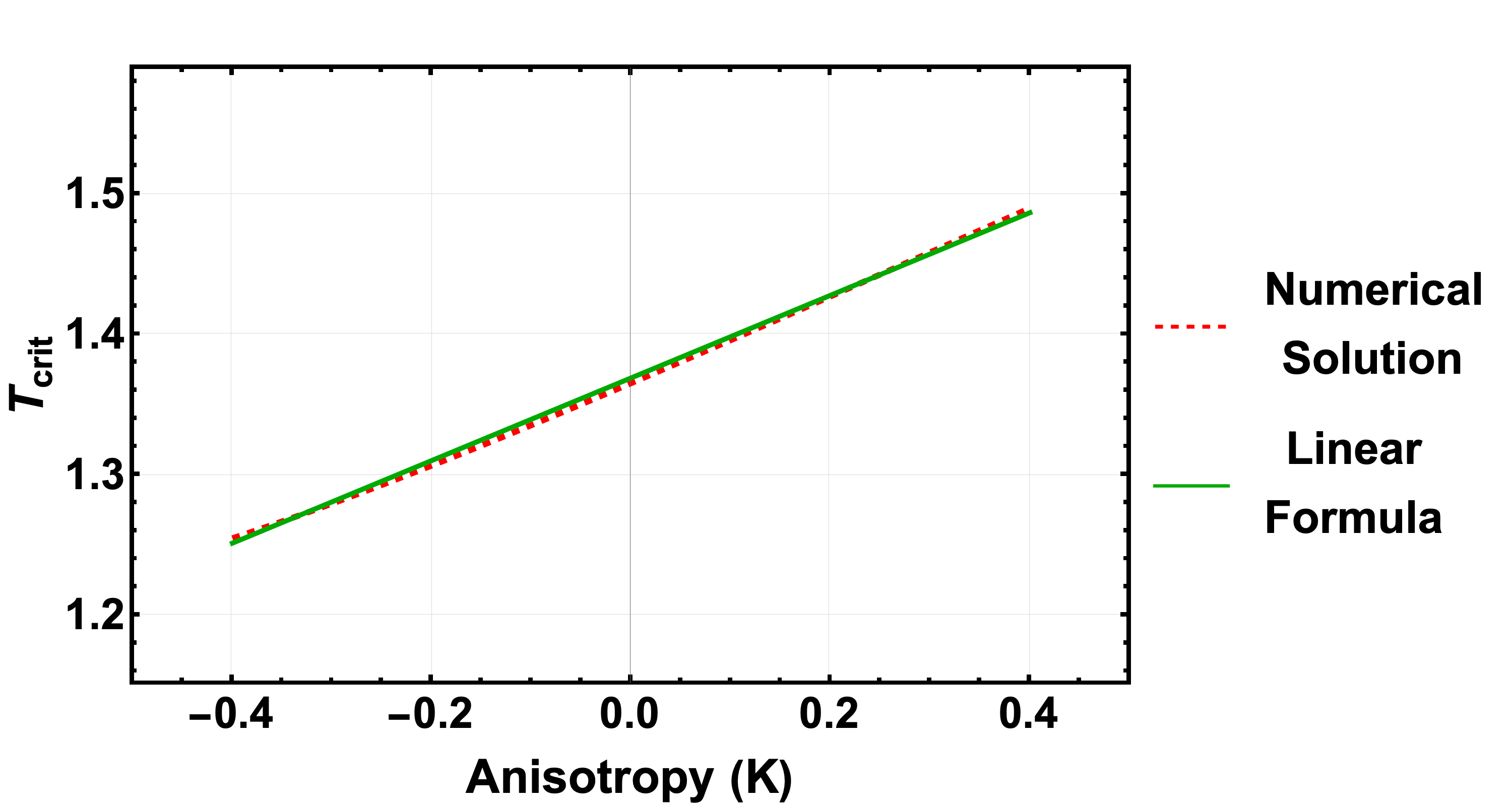}
    \caption{The magnetic anisotropy $K$ dependence of the critical temperature $T_{crit}$ between the antiferromagnetic $(J=1)$ ring $(\alpha=1)$ inverse magnetic susceptibility and $2T$. The numerical solutions to Eq.~\eqref{chiTcrit} (dashed red line) and the lineal Eq.~\ref{tcriteq} are depicted.}
    \label{Tcritic}
\end{figure}

{\subsection{Critical {Non-}Entanglement Temperature } 
 \begin{figure}
     \centering
     \includegraphics[width=.95\linewidth]{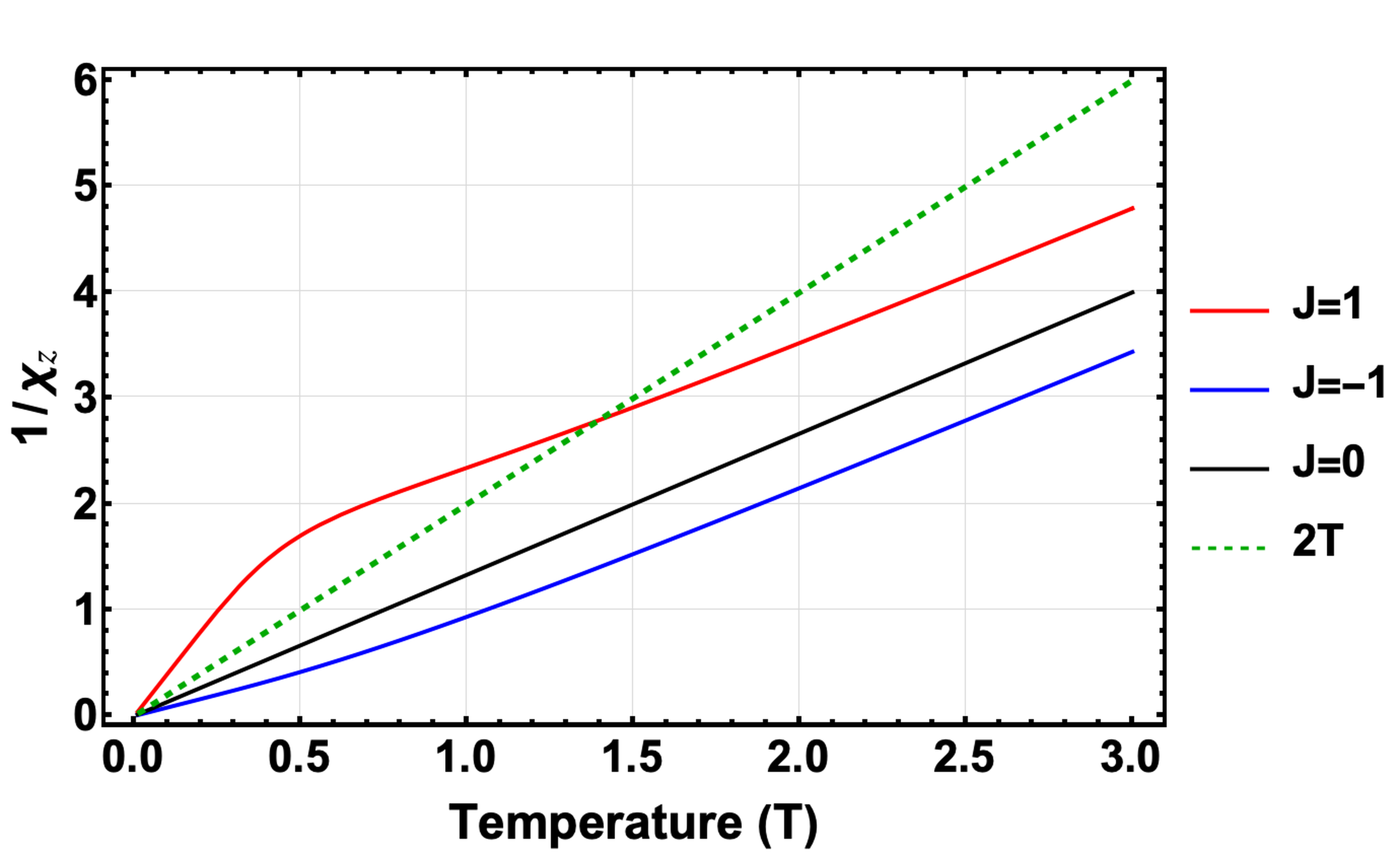}
     \caption{The temperature dependence of the ring ($\alpha=1$) inverse magnetic susceptibility $1/\chi_z$ at fixed magnetic anisotropy $K=0$ for different exchange interactions: Antiferromagnetic $J=1$ (solid red line); {ferromagnetic} $J=-1$ (solid blue line); Paramagnetic $J=0$ (solid black line). The green solid line represents the $2T$ minimum entanglement criterion for a three-qubit inverse susceptibility shown in Appendix \ref{Nqubitine} in Eq~\eqref{Nqubitineeq}. The lines $1/\chi_z$ below this green line indicate non-entanglement, while the lines above indicate entanglement. Only the antiferromagnetic inverse susceptibility crosses this line for a positive magnetic field.}
     \label{susceptability_temperature}
 \end{figure}

{As the antiferromagnetic system is entangled and magnetic anisotropy shifts the QCP,  the inverse susceptibility is consequently affected. Therefore, to describe the shift in the critical {temperature for the presence of entanglement} with anisotropy, we need to find the intersection between \(2T\) and \(1/\chi_z\), identifying the point where the system transitions from an entangled to a non-entangled system. Then, we have to solve the following equation using the relationship given by Eq.~\eqref{chiB0}:}
\eq{
\frac{1}{\chi_z}=2T_{\text{crit}}
\label{chiTcrit}
}
For the special case of \(K=0\), the crossing point can be solved analytically, given by \(T_{crit}(K=0) = \frac{3J}{2\ln{3}}\). However, due to the transcendental nature of solving Eq.~\eqref{chiTcrit}, we provide a linear formula for the crossing points in terms of {small} magnetic anisotropies {$(|K| \ll   1)$}:
\eq{
T_{crit} = 0.3K+ \frac{3}{2 \ln 3} \label{tcriteq}
}
The accuracy of this formula is demonstrated in Figure ~\ref{Tcritic}, where Eq.~\eqref{tcriteq} matches the numerical solutions of Eq.~\ref{chiTcrit}. This indicates that for higher {anisotropy {$K$} values, the critical temperature point increases,} suggesting that entanglement increases with higher anisotropy and decreases with lower anisotropy.
}

%%%%%%%%%%%%%%%%%%%%%%%%%%%%%%%%%%%%%%%%%%

%%%%%%%%%%%%%%%%%%%%%%%%%%%%%%%%%%%%%%%%%%
\section{Conclusions}
{We have conducted a comprehensive study of the Heisenberg \textit{XXX} model subjected to an external magnetic field along the \( z \)-axis, incorporating an anisotropy term in the \( y \)-direction. Our analysis focused on the behavior of an isolated qubit within this system, revealing the emergence of an effective temperature resulting from its interaction with the surrounding qubits. This temperature manifests in the reduced entropy of the system's ground state, obtained by tracing over the degrees of freedom of the other two qubits.}

{In the ferromagnetic regime, our results indicate that the effective temperature is highly sensitive to both the anisotropy constant and the external magnetic field. In contrast, in the antiferromagnetic ring regime---where the ground state does not depend on anisotropy---the system is susceptible to the initial weighting of the prepared states.

These findings underscore the pivotal role of anisotropy, ground-state weighting, and magnetic interactions in shaping the system's thermodynamic properties.}

Additionally, we computed the magnetic susceptibility across various regimes—antiferromagnetic, ferromagnetic, and paramagnetic. By establishing rigorous bounds on susceptibility for non-entangled states as a function of temperature, we identified the presence of entanglement when these bounds were violated.

Our findings show that entanglement occurs at low temperatures, specifically in the antiferromagnetic regime. Moreover, introducing anisotropy enhances the entanglement, extending the temperature range over which it persists. This enhancement underscores the intricate relationship between quantum interactions and thermodynamic properties, demonstrating the profound impact of anisotropy on the system's entanglement dynamics.

\section{Acknowledgments}
B.C, F.J.P., C.A, S.E.U. and P.V. acknowledge financial support from ANID
Fondecyt grant no. 1240582. B.C acknowledge PUCV and ''Direcci\'on de Postgrado'' of UTFSM.

%%%%%%%%%%%%%%%%%%%%%%%%%%%%%%%%%%%%%%%%%%
   %%%%%%%%%%%%%%%%%%%%%%%%%%%%%%%%%%%%%%%%%%
%% Optional

%% Only for journal Encyclopedia
%\entrylink{The Link_B T o this entry published on the encyclopedia platform.}

%%%%%%%%%%%%%%%%%%%%%%%%%%%%%%%%%%%%%%%%%%
%% Optional
%\appendixtitles{no} % Leave argument "no" if all appendix headings stay EMPTY (then no dot is printed after "Appendix A"). If the appendix sections contain a heading then change the argument to "yes".
%\appendixstart
\appendix

\section[\appendixname~\thesection]{Variable Definition}\label{variable_change}
The variables used in the article are, respectively
{\eq{P_\pm =\sqrt{B^2 \pm   BK +  K^2}  } 
\eq{R_{i}^\pm = \frac{1}{4}\qty( \pm 2B+ K +2(-1)^i  P_\pm )} 
\eq{L_{i}^\pm = \dfrac{1}{\displaystyle\sqrt{3+ \qty(\frac{ 4R_{i}^\pm }{K})^2}}}
  }

\eq{J_{\text{eff}}  = J (2+\alpha) + 5 K}
{\section[\appendixname~\thesection]{Non-entangled inequality for magnetic susceptibility}\label{Entanglement_Inequality}

We can construct a density matrix that is a direct tensor product of individual separable $\rho_i$ matrices such that
\eq{
\rho = \rho_1 \otimes \rho_2 \otimes \cdots \otimes \rho_n
}
where $\rho_i$ can be expressed as:
\eq{
\rho_i = p_i \ket{0}\bra{0} + (1-p_i)\ket{1}\bra{1}
}
Here, $p_i$ is a scalar between 0 and 1 associated with the state probability.

The total spin operators for the $n$ qubits can be expressed as:
{\eq{
\vu{S}_n &= \frac{1}{2} \sum_{\substack{\alpha = \\   \{x, y, z\}}}\sum_{k = 1}^n  \sigma_{\alpha k} }}
and the square of these operators is
{\eq{
\vu{S}_n^2 &= \frac{1}{4}  \sum_{kk'}^n \sum_{\alpha, \alpha'} \sigma_{\alpha k}\sigma_{\alpha' k'}
}}

The total magnetic susceptibility can be expressed in terms of magnetic fluctuations as:
\eq{
\chi = \frac{\expval{\vu{S}_n^2} - \expval{\vu{S}_n}^2}{k_B T}
}
where the mean expected value can be expressed as usual:
\eq{
\expval{\mathcal{O}} = \Tr{\rho \cdot \mathcal{O}}
}

\subsection{Two Qubits Example 1: product of mixed states}

For the case of two qubits, the density matrix has the form:
\eq{
\rho = \rho_1 \otimes \rho_2 \label{two_qubit_rho}
}
In matrix form:
\eq{
\rho &= \mqty(p_1 p_2 & 0 & 0 & 0 \\
0 & p_1(1-p_2) & 0 & 0 \\
0 & 0 & p_2(1-p_1) & 0 \\
0 & 0 & 0 & (1-p_1)(1-p_2)) \label{rho_2qubit_example_1}
}
The matrix form of the total spin operators for two qubits is:
\eq{
\vu{S}_2 &= {\frac{1}{2}}\mqty(2 & 1-i & 1-i & 0 \\
1+i & 0 & 0 & 1-i \\
1+i & 0 & 0 & 1-i \\
0 & 1+i & 1+i & -2) \label{S_2qubit}
}
Therefore, using the density matrix \eqref{rho_2qubit_example_1} with $\vu{S}_2$ \eqref{S_2qubit}, we have the expected value:
\eq{
\expval{\vu{S}_2}^2 =   (p_1 + p_2 - 1)^2
}
The square spin operator for two qubits $\vu{S}_2^2$ has the following matrix form:
\eq{
\vu{S}_2^2 &= {\frac{1}{2} }\left(\begin{array}{cccc}
4 & 1-i & 1-i & -2 i \\
1+i & 2 & 2 & -1+i \\
1+i & 2 & 2 & -1+i \\
2 i & -1-i & -1-i & 4
\end{array}\right)
}
Therefore, using the density matrix \eqref{rho_2qubit_example_1}, the expected value of this operator is:
\eq{
\expval{\vu{S}_2^2} = 2 - p_2 + p_1 (2p_2 - 1)
}
Consequently, the magnetic susceptibility is:
\eq{
\chi \cdot k_B T = 1+p_1+p_2- p_1^2- p_2^2
}

This function maximizes at 3/2 when \( p_1 \rightarrow \frac{1}{2} \) and \( p_2 \rightarrow \frac{1}{2} \) and minimizes at 1 when \( p_1 \rightarrow 1 \) and \( p_2 \rightarrow 1 \). This shows that the magnetic susceptibility for two qubits follows the inequality:

{
\eq{
 \frac{1}{k_B T} \leq \chi & \leq \frac{3}{2}\frac{1}{k_B T} \label{2qubitsxiinequality}
}
}

{
We must emphasize that this inequality and its lower bound are only valid for a product state, where each state is mixed. Previous reports have shown that \( \chi_z < 1 \) (in units of $k_{B}T$) can be observed in systems with high frustration, where competing interactions prevent uniform spin alignment, as seen in triangular lattice systems or spin-ice materials. Additionally, this behavior might arise in quantum spin liquids, where strong quantum fluctuations suppress magnetic ordering even at temperatures approaching absolute zero, resulting in diminished magnetic responses \cite{zhou2017quantum}. Furthermore, topological effects, such as those in quantum-protected states, could lead to unusually low magnetic susceptibilities due to long-range non-classical correlations \cite{mikitik2019magnetic}. Finally, dissipative systems or those governed by non-Hermitian Hamiltonians could exhibit suppressed magnetic susceptibility below the classical lower bound due to energy losses or environmental interactions \cite{hurst2022non,bergholtz2021exceptional}. These observations highlight the importance of further exploring scenarios where \( \chi_z < 1 \) might emerge and their potential connection to non-trivial quantum correlations.}

%Assuming that every spatial direction is equally probable, we have the special case for the \( z \) component of \(\chi\):
%\eq{
%\chi_z \geq \frac{1}{3 k_B T} \label{chi2qubitspatrion}
%}
{
\subsection{Two Qubits Example 2: product of Bloch states}
We consider the following product state
\begin{equation}
    \rho = \rho_1 \otimes \rho_2, 
\end{equation}
where each density matrix is is described by a pure state $\rho_i = \ket{\Psi_i}\bra{\Psi_i}$, where the following Bloch parametrization is used
\begin{equation}
    \ket{\Psi_i} = \cos\left({\theta_i \over 2} \right)\ket{0} + e^{i\phi_i} \sin\left( {\theta_i \over 2}\right)\ket{1}.
\end{equation}
Here, $\theta_i \in [0,\pi]$ and $\phi_i \in [0,2\pi]$. Using the above parameters, we found
\begin{eqnarray}
    \chi \cdot k_B T &=& 1 + {1 \over 8}\left[\cos(2\theta_1)\sin(2\phi_1)-\sin(2\phi_2) \right. \nonumber \\ 
    && \left.- \sin(2\phi_1) + \cos(2\theta_2)\sin(2\phi_2)) \right] \nonumber \\
    &&-{1 \over 4}\left[ \sin(2\theta_1)\cos(\phi_1) +\sin(2\theta_2)\cos(\phi_2)\right. \nonumber \\
    &&\left. \sin(2\theta_1)\sin(\phi_1) +\sin(2\theta_2)\sin(\phi_2) \right].
\end{eqnarray}
The minimum and maximum of the above function is reached when $\theta_1 = \theta_2$ and $\phi_1 = \phi_2$. By defining $\theta = \theta_1=\theta_2$ and $\phi = \phi_1 = \phi_2$, we get the surface
\begin{equation} \label{Surface_Chi}
    \chi \cdot k_B T = {1 \over 2}\left[3- \left[\cos \theta+\sin \theta(\cos \phi +\sin\phi) \right]^2\right].
\end{equation}
In Fig.~\ref{figura_Surface_Chi}, we show the behavior of the above surface, illustrating two important features. First, the susceptibility is constrained in the following manner

 \begin{figure}
     \centering
    \includegraphics[width=.95\linewidth]{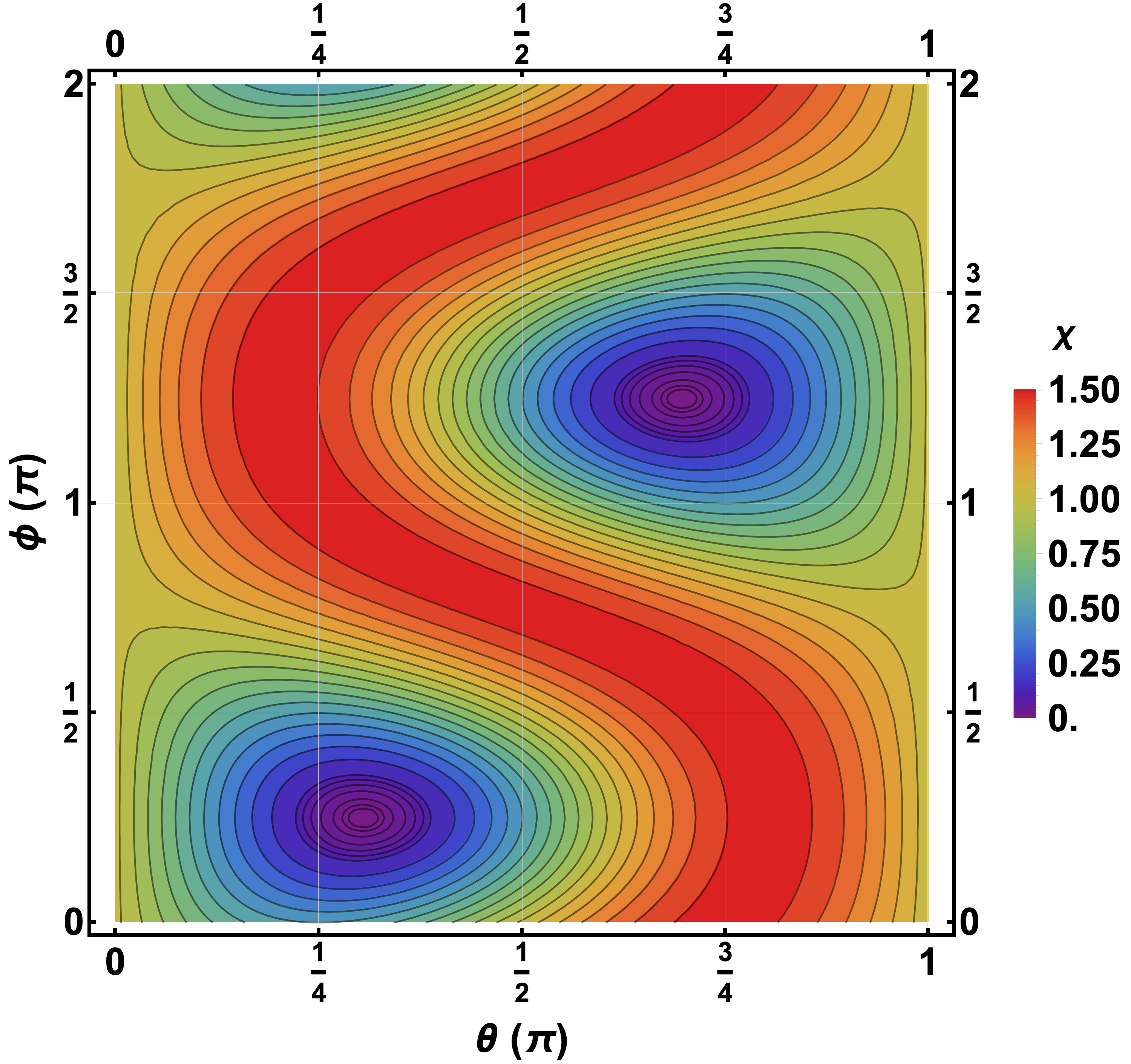}
     \caption{Surface $\chi \cdot k_B T$ as a function of $\theta$ and $\phi$ for the result shown in Eq.~\eqref{Surface_Chi}. The axis are in $\pi$ units.}
     \label{figura_Surface_Chi}
 \end{figure}
 
\begin{equation}
     0 \leq \chi \cdot k_B T \leq {3 \over 2},
\end{equation}

where the minimum susceptibility is equal to zero for two cases, namely, $(\theta, \phi) = (\pi/2-\tan^{-1}(2\sqrt{2})/2,\pi/4) \approx (0.30, 0.25)\pi$ and $(\pi/2+\tan^{-1}(2\sqrt{2})/2, 5\pi/4)\approx (0.70, 1.25)\pi$.

}

\subsection{Two Qubits: A Non-Separable Density Matrix}

Even though the system could be constructed using an entangled state, it will not violate necessarily the inequality Eq.~\eqref{2qubitsxiinequality}, as some entangled systems are more entangled than others. As an example of this we use the following Bell entangled state for two qubits:
\eq{
\psi_{Bell} = \cos \qty(\alpha) \ket{00} - \sin \qty(\alpha) \ket{11}
}
This state has a following density matrix:
\eq{
\rho_{Bell} &= \mqty(
\cos^2 \qty(\alpha) & 0 & 0 & -\cos \qty(\alpha) \sin \qty(\alpha) \\
0 & 0 & 0 & 0 \\
0 & 0 & 0 & 0 \\
-\cos \qty(\alpha) \sin \qty(\alpha) & 0 & 0 & \sin^2 \qty(\alpha) 
) \label{rho_bell_1}
}
For this case, using the density matrix \eqref{rho_bell_1}, the total spin operators are:
\eq{
\expval{\vu{S}_2}^2 =  \cos^2 \qty(2 \alpha)
}
and
\eq{
\expval{\vu{S}_2^2} = 2
}
Therefore, we can construct the magnetic susceptibility as:
\eq{
k_B T \chi = 2 - \cos^2 \qty(2 \alpha) \implies \chi \geq \frac{1}{k_B T}
}
It shares the same minimum as two non-entangled qubits in Eq.~\eqref{2qubitsxiinequality}. Thus, this measurement does not indicate whether the system is fully entangled or non-entangled.

\subsection{Adding a Singlet Density Matrix}
 {Our next candidate is the singlet state \(\ket{\Psi^-}\), a well-known two-spin state often discussed in the context of EPR-related experiments. This state is unique among the Bell states due to its phase difference. Like the other three Bell states \(\ket{\Psi^+}\), \(\ket{\Phi^+}\), and \(\ket{\Phi^-}\), it is maximally entangled, but it stands out because it consistently exhibits specific anticorrelations, independent of the measurement outcome.
We will start with the usual two-spin singlet state
}
{\eq{
\ket{\Psi^-} = \frac{1}{\sqrt{2}} \qty(\ket{10} - \ket{01}) \label{singlet_state}
}}
The singlet density matrix is:
\eq{
\rho_{singlet} &= \left(\begin{array}{cccc}
0 & 0 & 0 & 0 \\
0 & \frac{1}{2} & -\frac{1}{2} & 0 \\
0 & -\frac{1}{2} & \frac{1}{2} & 0 \\
0 & 0 & 0 & 0
\end{array}\right) \label{singlet_matrix}
}
If we add the singlet density matrix \eqref{singlet_matrix} to the two qubit non-separable density matrix \eqref{rho_bell_1} with probabilities \( p \) and \( (1-p) \) respectively, we have that the spin operators' expected values for this system are:
\eq{
\rho &= p\rho_{Bell} + (1-p)\rho_{singlet} \\
&= \mqty(p \cos^2 \qty(\alpha) & 0 & 0 & -p \cos \qty(\alpha) \sin \qty(\alpha) \\
0 & \frac{1-p}{2} & \frac{p-1}{2} & 0 \\
0 & \frac{p-1}{2} & \frac{1-p}{2} & 0 \\
- p \cos \qty(\alpha) \sin \qty(\alpha) & 0 & 0 & p \sin^2 \qty(\alpha)
)
}
The square spin operator is:
\eq{
\expval{\vu{S}_2}^2 =  p^2 \cos^2 \qty(2 \alpha)
}
and:
\eq{
\expval{\vu{S}_2^2} = 2p
}
Therefore, the magnetic susceptibility of the mixed system between the singlet state and the other Bell state is:
\eq{
k_B T \chi = p (2- p \cos^2 \qty(2\alpha))
}
{We observe that the minimum value is zero when \( p = 0 \). This occurs the system is maximally entangled in the singlet state, violating the inequality \eqref{2qubitsxiinequality}. This finding illustrates that a classical thermodynamic quantity serves as an entanglement witness.}

{\subsection{Transforming the Singlet into the other Bell states}
Given that the Bell states can be transformed into one another through local operations \cite{Bell1}, we multiply the singlet state \(\ket{\Psi^-}\) \eqref{singlet_state} by the Pauli matrices to test each transformation. To ensure that the entanglement witness consistently detects entanglement across different state configurations, it will be reevaluated for each Bell state.

The remaining Bell states are defined as follows:
\eq{
\ket{\Psi^+} = \sigma_z^{(2)} \ket{\Psi^-} = \frac{1}{\sqrt{2}} \qty(\ket{10} + \ket{01})
\\
\ket{\Phi^-} = \sigma_x^{(1)} \ket{\Psi^-} = \frac{1}{\sqrt{2}} \qty(\ket{00} - \ket{11})\\
\ket{\Phi^+} = \sigma_x^{(1)}\sigma_z^{(2)} \ket{\Psi^-} = \frac{1}{\sqrt{2}} \qty(\ket{00} + \ket{11})
} 
Adding their associated density matrices to the non-separable density matrix \eqref{rho_bell_1} with probabilities $p_i$ and $(1-p_i)$ results in:
\eq{
\rho(i) &= p_i \rho_{Bell} + (1-p_i)\rho_{i}}
The specific density matrices for each Bell state are:
\eq{ 
\rescale[.5]{\rho ({{\Psi^+}} )= \begin{pmatrix}
p \cos^2(\alpha) & 0 & 0 & -p \cos(\alpha) \sin(\alpha) \\
0 & \frac{1 - p}{2} & \frac{1 - p}{2} & 0 \\
0 & \frac{1 - p}{2} & \frac{1 - p}{2} & 0 \\
-p \cos(\alpha) \sin(\alpha) & 0 & 0 & p \sin^2(\alpha)
\end{pmatrix}} \label{rhobell1}
}
\eq{ 
\rescale[.5]{\rho ({{\Phi^-}} )= \frac{1}{2}\begin{pmatrix}
 \left( 1 + p_{{{\Phi^-}}} \cos(2\alpha) \right) & 0 & 0 &  \left( -1 + p_{{\Phi^-}}- 2p_{{\Phi^-}} \cos(\alpha) \sin(\alpha) \right) \\
0 & 0 & 0 & 0 \\
0 & 0 & 0 & 0 \\
 \left( -1 + p{{\Phi^-}} - 2 p_{{\Phi^-}}   \cos(\alpha) \sin(\alpha) \right) & 0 & 0 &   \left( 1 - p_{{\Phi^-}} \cos(2\alpha) \right)
\end{pmatrix}} \label{rhobell2}
}
\eq{ 
\rescale[.5]{\rho ({{\Phi^+}}) = \frac{1}{2}\begin{pmatrix}
 \left( 1 + p_{{\Phi^+}} \cos(2\alpha) \right) & 0 & 0 &  \left( 1 - p_{{\Phi^+}} - 2p_{{\Phi^+}} \cos(\alpha) \sin(\alpha) \right) \\
0 & 0 & 0 & 0 \\
0 & 0 & 0 & 0 \\
 \left( 1 - p_{{\Phi^+}} - 2 p_{{\Phi^+}} \cos(\alpha) \sin(\alpha) \right) & 0 & 0 &   \left( 1 - p_{{\Phi^+}} \cos(2\alpha) \right)
\end{pmatrix}} \label{rhobell3}
}
Although the density matrices \eqref{rhobell1}, \eqref{rhobell2}, and \eqref{rhobell3} differs each other, the expected values of the spin operators are identical for all the three associated states: \eq{\expval{\vu{S}_2}^2 &= p^2 \cos^2\qty(2 \alpha) && \expval{\vu{S}_2^2} = 2}
With magnetic susceptibility having the same minimum as two separable qubits
\eq{
k_B T \chi  = 2 - p^2 \cos^2 (2 \alpha) \xrightarrow{p \rightarrow 1} \chi_{\text{min}} = \frac{1}{k_B T }
}
The entanglement witness is not effective at detecting entanglement in the three maximally entangled Bell states, which exhibit symmetric correlations, distinguishing them from the singlet state that displays specific anti-correlations and a unique phase difference. It can only identify entanglement in specific scenarios, such as when their correlations are negative, as demonstrated in Eq.~\eqref{Nqubitineeq1}.

\subsection{Two Qubits: Werner state }

The convex set of Werner states \cite{Werner1,Werner2} can be represented as a combination of the totally mixed state and the projection onto a singlet state, given by:
\begin{equation}
W(\lambda) = \lambda \left|\Psi^{-}\right\rangle \left\langle\Psi^{-}\right| + \frac{1-\lambda}{4} I_{4 \times 4},
\end{equation}
where $\lambda \in [0,1]$. The density matrix of the Werner state is expressed as:
\eq{
\rho_{W}(\lambda)= \frac{1}{4} \left(\begin{array}{cccc}
1-\lambda & 0 & 0 & 0 \\
0 & 1+\lambda & -2 \lambda & 0 \\
0 & -2 \lambda & 1+\lambda & 0 \\
0 & 0 & 0 & 1-\lambda
\end{array}\right) \label{wigner_rho}
} 

Using this density matrix \eqref{wigner_rho}, the total spin operators are computed as follows:
\begin{align}
\expval{\vu{S}_2}^2 &= 0, & \expval{\vu{S}_2^2} &= \frac{3}{2}(1-\lambda).
\end{align}

Correspondingly, the magnetic susceptibility of the Werner density matrix is:
\begin{equation}
\chi_W = \frac{3}{2k_B T}(1-\lambda).
\end{equation}

It has been shown in the literature that the system is entangled for $\lambda > 1/3$. At $\lambda = 1/3$, the system exhibits the minimum magnetic susceptibility for two mixed states \eqref{2qubitsxiinequality}. For $\lambda > 1/3$, the inequality is violated, clearly demonstrating the presence of entanglement. However, in the entangled regime, $\lambda \in [1/3, 1]$, this thermodynamic quantity cannot distinguish how strongly or weakly entangled the system is.

}

\subsection{Three qubits example}

For three qubits non-entangled density matrix:
\eq{
\rho = \rho_1 \otimes \rho_2 \otimes \rho_3 \label{rho_3qubit_example_1}
}
we will omit the explicit matrix form for the sake of space. The spin operators can be constructed similarly, and using the density matrix \eqref{rho_3qubit_example_1}, we can obtain their expected value:
\eq{
\expval{\vu{S}_3}^2 =  \qty(p_1+p_2+p_3-\frac{3}{2})^2
}
The expected value of the square of the total spin operator has the form:
\eq{
\expval{\vu{S}_3^2} = \frac{15}{4} +  2p_2 (p_3-1)+2p_1(p_2+p_3-1)-2p_3
}
Therefore, the magnetic susceptibility for three separable qubits is:
\eq{
k_B T \chi = \frac{3}{2} - p_1(p_1 - 1) -p_2(p_2 - 1) - p_3(p_3 - 1)
}
where its minimum is  $\frac{3}{2}$ when \( p_1, p_2, p_3 \rightarrow 1 \) and its maximum occurs at $9/4$ when \( p_1, p_2, p_3 \rightarrow \frac{1}{2} \).\\
This shows the following inequality for the \( z \) component of the three qubits:
\eq{
\chi_z \geq \frac{1}{2k_B T} \implies \frac{1}{\chi_z }\leq  {2k_B T}\label{3qubitsxiinequality}
}

Thus, if an experimental measurement yields a value that does not follow this inequality, it indicates the presence of entanglement in the system.

\subsection{$N$ qubit inequality}\label{Nqubitine}

In the literature \cite{WOS:000234347200003}, using a different method, the generalization of this inequality for \( N \) qubits has been formally demonstrated such as:
\eq{
\chi \equiv \chi_x + \chi_y + \chi_z \geqslant \frac{N s}{k_B T }, \label{Nqubitintotaleq}
}
where $N$ is the number of qubits, and $s$ is the spin. Therefore, the entanglement criterion is:
\eq{
\chi_z < \frac{1}{3} \frac{N s}{k_B T} \label{Nqubitineeq}
}
Our result for $s=1/2$, using matrices and minimizing probability variables, can be generalized to obtain the same result, but with the disadvantage of eventually having to formally minimize \( N \) different variables, which will require more computational time. However, if we extrapolate the pattern obtained from two qubits \eqref{2qubitsxiinequality} and three qubits \eqref{3qubitsxiinequality}, it will provide us with the same value for the \( N \)-th magnetic susceptibility \eqref{Nqubitintotaleq} {for spin particles}.

 %\externalbibliography{yes}
%\input{biblio_3qubit.bbl}
% Reference 1

%\bibliography{biblio_3qubit.bib}

\begin{thebibliography}{52}
\expandafter\ifx\csname natexlab\endcsname\relax\def\natexlab#1{#1}\fi
\expandafter\ifx\csname bibnamefont\endcsname\relax
  \def\bibnamefont#1{#1}\fi
\expandafter\ifx\csname bibfnamefont\endcsname\relax
  \def\bibfnamefont#1{#1}\fi
\expandafter\ifx\csname citenamefont\endcsname\relax
  \def\citenamefont#1{#1}\fi
\expandafter\ifx\csname url\endcsname\relax
  \def\url#1{\texttt{#1}}\fi
\expandafter\ifx\csname urlprefix\endcsname\relax\def\urlprefix{URL }\fi
\providecommand{\bibinfo}[2]{#2}
\providecommand{\eprint}[2][]{\url{#2}}

\bibitem[{\citenamefont{Gomez-Coca and Ruiz}(2013)}]{WOS:000324304400012}
\bibinfo{author}{\bibfnamefont{S.}~\bibnamefont{Gomez-Coca}} \bibnamefont{and} \bibinfo{author}{\bibfnamefont{E.}~\bibnamefont{Ruiz}}, \bibinfo{journal}{CANADIAN JOURNAL OF CHEMISTRY} \textbf{\bibinfo{volume}{91}}, \bibinfo{pages}{866} (\bibinfo{year}{2013}), ISSN \bibinfo{issn}{0008-4042}.

\bibitem[{\citenamefont{Chiesa et~al.}(2024)\citenamefont{Chiesa, Santini, Garlatti, Luis, and Carretta}}]{WOS:001156816000001}
\bibinfo{author}{\bibfnamefont{A.}~\bibnamefont{Chiesa}}, \bibinfo{author}{\bibfnamefont{P.}~\bibnamefont{Santini}}, \bibinfo{author}{\bibfnamefont{E.}~\bibnamefont{Garlatti}}, \bibinfo{author}{\bibfnamefont{F.}~\bibnamefont{Luis}}, \bibnamefont{and} \bibinfo{author}{\bibfnamefont{S.}~\bibnamefont{Carretta}}, \bibinfo{journal}{REPORTS ON PROGRESS IN PHYSICS} \textbf{\bibinfo{volume}{87}} (\bibinfo{year}{2024}), ISSN \bibinfo{issn}{0034-4885}.

\bibitem[{\citenamefont{Fursina and Sinitskii}(2023)}]{WOS:001027763900001}
\bibinfo{author}{\bibfnamefont{A.~A.} \bibnamefont{Fursina}} \bibnamefont{and} \bibinfo{author}{\bibfnamefont{A.}~\bibnamefont{Sinitskii}}, \bibinfo{journal}{ACS APPLIED ELECTRONIC MATERIALS} \textbf{\bibinfo{volume}{5}}, \bibinfo{pages}{3531} (\bibinfo{year}{2023}).

\bibitem[{\citenamefont{Smith et~al.}(2020)\citenamefont{Smith, Wysocki, and Park}}]{sergio1}
\bibinfo{author}{\bibfnamefont{R.~L.} \bibnamefont{Smith}}, \bibinfo{author}{\bibfnamefont{A.~L.} \bibnamefont{Wysocki}}, \bibnamefont{and} \bibinfo{author}{\bibfnamefont{K.}~\bibnamefont{Park}}, \bibinfo{journal}{PHYSICAL CHEMISTRY CHEMICAL PHYSICS} \textbf{\bibinfo{volume}{22}}, \bibinfo{pages}{21793} (\bibinfo{year}{2020}), ISSN \bibinfo{issn}{1463-9076}.

\bibitem[{\citenamefont{Goss et~al.}(2022)\citenamefont{Goss, Morvan, Marinelli, Mitchell, Nguyen, Naik, Chen, Junger, Kreikebaum, Santiago et~al.}}]{WOS:000952021200013}
\bibinfo{author}{\bibfnamefont{N.}~\bibnamefont{Goss}}, \bibinfo{author}{\bibfnamefont{A.}~\bibnamefont{Morvan}}, \bibinfo{author}{\bibfnamefont{B.}~\bibnamefont{Marinelli}}, \bibinfo{author}{\bibfnamefont{B.~K.} \bibnamefont{Mitchell}}, \bibinfo{author}{\bibfnamefont{L.~B.} \bibnamefont{Nguyen}}, \bibinfo{author}{\bibfnamefont{R.~K.} \bibnamefont{Naik}}, \bibinfo{author}{\bibfnamefont{L.}~\bibnamefont{Chen}}, \bibinfo{author}{\bibfnamefont{C.}~\bibnamefont{Junger}}, \bibinfo{author}{\bibfnamefont{J.~M.} \bibnamefont{Kreikebaum}}, \bibinfo{author}{\bibfnamefont{D.~I.} \bibnamefont{Santiago}}, \bibnamefont{et~al.}, \bibinfo{journal}{NATURE COMMUNICATIONS} \textbf{\bibinfo{volume}{13}} (\bibinfo{year}{2022}).

\bibitem[{\citenamefont{Arute et~al.}(2019)\citenamefont{Arute, Arya, Babbush, Bacon, Bardin, Barends, Biswas, Boixo, Brandao, Buell et~al.}}]{WOS:000492991700045}
\bibinfo{author}{\bibfnamefont{F.}~\bibnamefont{Arute}}, \bibinfo{author}{\bibfnamefont{K.}~\bibnamefont{Arya}}, \bibinfo{author}{\bibfnamefont{R.}~\bibnamefont{Babbush}}, \bibinfo{author}{\bibfnamefont{D.}~\bibnamefont{Bacon}}, \bibinfo{author}{\bibfnamefont{J.~C.} \bibnamefont{Bardin}}, \bibinfo{author}{\bibfnamefont{R.}~\bibnamefont{Barends}}, \bibinfo{author}{\bibfnamefont{R.}~\bibnamefont{Biswas}}, \bibinfo{author}{\bibfnamefont{S.}~\bibnamefont{Boixo}}, \bibinfo{author}{\bibfnamefont{F.~G. S.~L.} \bibnamefont{Brandao}}, \bibinfo{author}{\bibfnamefont{D.~A.} \bibnamefont{Buell}}, \bibnamefont{et~al.}, \bibinfo{journal}{NATURE} \textbf{\bibinfo{volume}{574}}, \bibinfo{pages}{505+} (\bibinfo{year}{2019}), ISSN \bibinfo{issn}{0028-0836}.

\bibitem[{\citenamefont{Hempel et~al.}(2018)\citenamefont{Hempel, Maier, Romero, McClean, Monz, Shen, Jurcevic, Lanyon, Love, Babbush et~al.}}]{WOS:000439545700001}
\bibinfo{author}{\bibfnamefont{C.}~\bibnamefont{Hempel}}, \bibinfo{author}{\bibfnamefont{C.}~\bibnamefont{Maier}}, \bibinfo{author}{\bibfnamefont{J.}~\bibnamefont{Romero}}, \bibinfo{author}{\bibfnamefont{J.}~\bibnamefont{McClean}}, \bibinfo{author}{\bibfnamefont{T.}~\bibnamefont{Monz}}, \bibinfo{author}{\bibfnamefont{H.}~\bibnamefont{Shen}}, \bibinfo{author}{\bibfnamefont{P.}~\bibnamefont{Jurcevic}}, \bibinfo{author}{\bibfnamefont{B.~P.} \bibnamefont{Lanyon}}, \bibinfo{author}{\bibfnamefont{P.}~\bibnamefont{Love}}, \bibinfo{author}{\bibfnamefont{R.}~\bibnamefont{Babbush}}, \bibnamefont{et~al.}, \bibinfo{journal}{PHYSICAL REVIEW X} \textbf{\bibinfo{volume}{8}} (\bibinfo{year}{2018}), ISSN \bibinfo{issn}{2160-3308}.

\bibitem[{\citenamefont{Friis et~al.}(2018)\citenamefont{Friis, Marty, Maier, Hempel, Holzaepfel, Jurcevic, Plenio, Huber, Roos, Blatt et~al.}}]{WOS:000429529600002}
\bibinfo{author}{\bibfnamefont{N.}~\bibnamefont{Friis}}, \bibinfo{author}{\bibfnamefont{O.}~\bibnamefont{Marty}}, \bibinfo{author}{\bibfnamefont{C.}~\bibnamefont{Maier}}, \bibinfo{author}{\bibfnamefont{C.}~\bibnamefont{Hempel}}, \bibinfo{author}{\bibfnamefont{M.}~\bibnamefont{Holzaepfel}}, \bibinfo{author}{\bibfnamefont{P.}~\bibnamefont{Jurcevic}}, \bibinfo{author}{\bibfnamefont{M.~B.} \bibnamefont{Plenio}}, \bibinfo{author}{\bibfnamefont{M.}~\bibnamefont{Huber}}, \bibinfo{author}{\bibfnamefont{C.}~\bibnamefont{Roos}}, \bibinfo{author}{\bibfnamefont{R.}~\bibnamefont{Blatt}}, \bibnamefont{et~al.}, \bibinfo{journal}{PHYSICAL REVIEW X} \textbf{\bibinfo{volume}{8}} (\bibinfo{year}{2018}), ISSN \bibinfo{issn}{2160-3308}.

\bibitem[{\citenamefont{Zhang et~al.}(2017)\citenamefont{Zhang, Pagano, Hess, Kyprianidis, Ecker, Kaplan, Gorshkov, Gong, and Monroe}}]{WOS:000416520400037}
\bibinfo{author}{\bibfnamefont{J.}~\bibnamefont{Zhang}}, \bibinfo{author}{\bibfnamefont{G.}~\bibnamefont{Pagano}}, \bibinfo{author}{\bibfnamefont{P.~W.} \bibnamefont{Hess}}, \bibinfo{author}{\bibfnamefont{A.}~\bibnamefont{Kyprianidis}}, \bibinfo{author}{\bibfnamefont{P.~B.} \bibnamefont{Ecker}}, \bibinfo{author}{\bibfnamefont{H.}~\bibnamefont{Kaplan}}, \bibinfo{author}{\bibfnamefont{A.~V.} \bibnamefont{Gorshkov}}, \bibinfo{author}{\bibfnamefont{Z.~X.} \bibnamefont{Gong}}, \bibnamefont{and} \bibinfo{author}{\bibfnamefont{C.}~\bibnamefont{Monroe}}, \bibinfo{journal}{NATURE} \textbf{\bibinfo{volume}{551}}, \bibinfo{pages}{601+} (\bibinfo{year}{2017}), ISSN \bibinfo{issn}{0028-0836}.

\bibitem[{\citenamefont{Ringbauer et~al.}(2022)\citenamefont{Ringbauer, Meth, Postler, Stricker, Blatt, Schindler, and Monz}}]{WOS:000828441600002}
\bibinfo{author}{\bibfnamefont{M.}~\bibnamefont{Ringbauer}}, \bibinfo{author}{\bibfnamefont{M.}~\bibnamefont{Meth}}, \bibinfo{author}{\bibfnamefont{L.}~\bibnamefont{Postler}}, \bibinfo{author}{\bibfnamefont{R.}~\bibnamefont{Stricker}}, \bibinfo{author}{\bibfnamefont{R.}~\bibnamefont{Blatt}}, \bibinfo{author}{\bibfnamefont{P.}~\bibnamefont{Schindler}}, \bibnamefont{and} \bibinfo{author}{\bibfnamefont{T.}~\bibnamefont{Monz}}, \bibinfo{journal}{NATURE PHYSICS} \textbf{\bibinfo{volume}{18}}, \bibinfo{pages}{1053+} (\bibinfo{year}{2022}), ISSN \bibinfo{issn}{1745-2473}.

\bibitem[{\citenamefont{Madsen et~al.}(2022)\citenamefont{Madsen, Laudenbach, Askarani, Rortais, Vincent, Bulmer, Miatto, Neuhaus, Helt, Collins et~al.}}]{WOS:000804867100011}
\bibinfo{author}{\bibfnamefont{L.~S.} \bibnamefont{Madsen}}, \bibinfo{author}{\bibfnamefont{F.}~\bibnamefont{Laudenbach}}, \bibinfo{author}{\bibfnamefont{M.~F.} \bibnamefont{Askarani}}, \bibinfo{author}{\bibfnamefont{F.}~\bibnamefont{Rortais}}, \bibinfo{author}{\bibfnamefont{T.}~\bibnamefont{Vincent}}, \bibinfo{author}{\bibfnamefont{J.~F.~F.} \bibnamefont{Bulmer}}, \bibinfo{author}{\bibfnamefont{F.~M.} \bibnamefont{Miatto}}, \bibinfo{author}{\bibfnamefont{L.}~\bibnamefont{Neuhaus}}, \bibinfo{author}{\bibfnamefont{L.~G.} \bibnamefont{Helt}}, \bibinfo{author}{\bibfnamefont{M.~J.} \bibnamefont{Collins}}, \bibnamefont{et~al.}, \bibinfo{journal}{NATURE} \textbf{\bibinfo{volume}{606}}, \bibinfo{pages}{75+} (\bibinfo{year}{2022}), ISSN \bibinfo{issn}{0028-0836}.

\bibitem[{\citenamefont{Zhong et~al.}(2020)\citenamefont{Zhong, Wang, Deng, Chen, Peng, Luo, Qin, Wu, Ding, Hu et~al.}}]{WOS:000599959400045}
\bibinfo{author}{\bibfnamefont{H.-S.} \bibnamefont{Zhong}}, \bibinfo{author}{\bibfnamefont{H.}~\bibnamefont{Wang}}, \bibinfo{author}{\bibfnamefont{Y.-H.} \bibnamefont{Deng}}, \bibinfo{author}{\bibfnamefont{M.-C.} \bibnamefont{Chen}}, \bibinfo{author}{\bibfnamefont{L.-C.} \bibnamefont{Peng}}, \bibinfo{author}{\bibfnamefont{Y.-H.} \bibnamefont{Luo}}, \bibinfo{author}{\bibfnamefont{J.}~\bibnamefont{Qin}}, \bibinfo{author}{\bibfnamefont{D.}~\bibnamefont{Wu}}, \bibinfo{author}{\bibfnamefont{X.}~\bibnamefont{Ding}}, \bibinfo{author}{\bibfnamefont{Y.}~\bibnamefont{Hu}}, \bibnamefont{et~al.}, \bibinfo{journal}{SCIENCE} \textbf{\bibinfo{volume}{370}}, \bibinfo{pages}{1460} (\bibinfo{year}{2020}), ISSN \bibinfo{issn}{0036-8075}.

\bibitem[{\citenamefont{Noiri et~al.}(2022)\citenamefont{Noiri, Takeda, Nakajima, Kobayashi, Sammak, Scappucci, and Tarucha}}]{WOS:000744666700023}
\bibinfo{author}{\bibfnamefont{A.}~\bibnamefont{Noiri}}, \bibinfo{author}{\bibfnamefont{K.}~\bibnamefont{Takeda}}, \bibinfo{author}{\bibfnamefont{T.}~\bibnamefont{Nakajima}}, \bibinfo{author}{\bibfnamefont{T.}~\bibnamefont{Kobayashi}}, \bibinfo{author}{\bibfnamefont{A.}~\bibnamefont{Sammak}}, \bibinfo{author}{\bibfnamefont{G.}~\bibnamefont{Scappucci}}, \bibnamefont{and} \bibinfo{author}{\bibfnamefont{S.}~\bibnamefont{Tarucha}}, \bibinfo{journal}{NATURE} \textbf{\bibinfo{volume}{601}}, \bibinfo{pages}{338+} (\bibinfo{year}{2022}), ISSN \bibinfo{issn}{0028-0836}.

\bibitem[{\citenamefont{Madzik et~al.}(2022)\citenamefont{Madzik, Asaad, Youssry, Joecker, Rudinger, Nielsen, Young, Proctor, Baczewski, Laucht et~al.}}]{WOS:000744666700025}
\bibinfo{author}{\bibfnamefont{M.~T.} \bibnamefont{Madzik}}, \bibinfo{author}{\bibfnamefont{S.}~\bibnamefont{Asaad}}, \bibinfo{author}{\bibfnamefont{A.}~\bibnamefont{Youssry}}, \bibinfo{author}{\bibfnamefont{B.}~\bibnamefont{Joecker}}, \bibinfo{author}{\bibfnamefont{K.~M.} \bibnamefont{Rudinger}}, \bibinfo{author}{\bibfnamefont{E.}~\bibnamefont{Nielsen}}, \bibinfo{author}{\bibfnamefont{K.~C.} \bibnamefont{Young}}, \bibinfo{author}{\bibfnamefont{T.~J.} \bibnamefont{Proctor}}, \bibinfo{author}{\bibfnamefont{A.~D.} \bibnamefont{Baczewski}}, \bibinfo{author}{\bibfnamefont{A.}~\bibnamefont{Laucht}}, \bibnamefont{et~al.}, \bibinfo{journal}{NATURE} \textbf{\bibinfo{volume}{601}}, \bibinfo{pages}{348+} (\bibinfo{year}{2022}), ISSN \bibinfo{issn}{0028-0836}.

\bibitem[{\citenamefont{Xue et~al.}(2022)\citenamefont{Xue, Russ, Samkharadze, Undseth, Sammak, Scappucci, and Vandersypen}}]{WOS:000744666700024}
\bibinfo{author}{\bibfnamefont{X.}~\bibnamefont{Xue}}, \bibinfo{author}{\bibfnamefont{M.}~\bibnamefont{Russ}}, \bibinfo{author}{\bibfnamefont{N.}~\bibnamefont{Samkharadze}}, \bibinfo{author}{\bibfnamefont{B.}~\bibnamefont{Undseth}}, \bibinfo{author}{\bibfnamefont{A.}~\bibnamefont{Sammak}}, \bibinfo{author}{\bibfnamefont{G.}~\bibnamefont{Scappucci}}, \bibnamefont{and} \bibinfo{author}{\bibfnamefont{L.~M.~K.} \bibnamefont{Vandersypen}}, \bibinfo{journal}{NATURE} \textbf{\bibinfo{volume}{601}}, \bibinfo{pages}{343+} (\bibinfo{year}{2022}), ISSN \bibinfo{issn}{0028-0836}.

\bibitem[{\citenamefont{Arnesen et~al.}(2001)\citenamefont{Arnesen, Bose, and Vedral}}]{WOS:000169822900048}
\bibinfo{author}{\bibfnamefont{M.}~\bibnamefont{Arnesen}}, \bibinfo{author}{\bibfnamefont{S.}~\bibnamefont{Bose}}, \bibnamefont{and} \bibinfo{author}{\bibfnamefont{V.}~\bibnamefont{Vedral}}, \bibinfo{journal}{PHYSICAL REVIEW LETTERS} \textbf{\bibinfo{volume}{87}} (\bibinfo{year}{2001}), ISSN \bibinfo{issn}{0031-9007}.

\bibitem[{\citenamefont{Maurer et~al.}(2012)\citenamefont{Maurer, Kucsko, Latta, Jiang, Yao, Bennett, Pastawski, Hunger, Chisholm, Markham et~al.}}]{WOS:000304905300040}
\bibinfo{author}{\bibfnamefont{P.~C.} \bibnamefont{Maurer}}, \bibinfo{author}{\bibfnamefont{G.}~\bibnamefont{Kucsko}}, \bibinfo{author}{\bibfnamefont{C.}~\bibnamefont{Latta}}, \bibinfo{author}{\bibfnamefont{L.}~\bibnamefont{Jiang}}, \bibinfo{author}{\bibfnamefont{N.~Y.} \bibnamefont{Yao}}, \bibinfo{author}{\bibfnamefont{S.~D.} \bibnamefont{Bennett}}, \bibinfo{author}{\bibfnamefont{F.}~\bibnamefont{Pastawski}}, \bibinfo{author}{\bibfnamefont{D.}~\bibnamefont{Hunger}}, \bibinfo{author}{\bibfnamefont{N.}~\bibnamefont{Chisholm}}, \bibinfo{author}{\bibfnamefont{M.}~\bibnamefont{Markham}}, \bibnamefont{et~al.}, \bibinfo{journal}{SCIENCE} \textbf{\bibinfo{volume}{336}}, \bibinfo{pages}{1283} (\bibinfo{year}{2012}), ISSN \bibinfo{issn}{0036-8075}.

\bibitem[{\citenamefont{Neumann et~al.}(2008)\citenamefont{Neumann, Mizuochi, Rempp, Hemmer, Watanabe, Yamasaki, Jacques, Gaebel, Jelezko, and Wrachtrup}}]{WOS:000256441100041}
\bibinfo{author}{\bibfnamefont{P.}~\bibnamefont{Neumann}}, \bibinfo{author}{\bibfnamefont{N.}~\bibnamefont{Mizuochi}}, \bibinfo{author}{\bibfnamefont{F.}~\bibnamefont{Rempp}}, \bibinfo{author}{\bibfnamefont{P.}~\bibnamefont{Hemmer}}, \bibinfo{author}{\bibfnamefont{H.}~\bibnamefont{Watanabe}}, \bibinfo{author}{\bibfnamefont{S.}~\bibnamefont{Yamasaki}}, \bibinfo{author}{\bibfnamefont{V.}~\bibnamefont{Jacques}}, \bibinfo{author}{\bibfnamefont{T.}~\bibnamefont{Gaebel}}, \bibinfo{author}{\bibfnamefont{F.}~\bibnamefont{Jelezko}}, \bibnamefont{and} \bibinfo{author}{\bibfnamefont{J.}~\bibnamefont{Wrachtrup}}, \bibinfo{journal}{SCIENCE} \textbf{\bibinfo{volume}{320}}, \bibinfo{pages}{1326} (\bibinfo{year}{2008}), ISSN \bibinfo{issn}{0036-8075}.

\bibitem[{\citenamefont{Awschalom et~al.}(2013)\citenamefont{Awschalom, Bassett, Dzurak, Hu, and Petta}}]{WOS:000315709900039}
\bibinfo{author}{\bibfnamefont{D.~D.} \bibnamefont{Awschalom}}, \bibinfo{author}{\bibfnamefont{L.~C.} \bibnamefont{Bassett}}, \bibinfo{author}{\bibfnamefont{A.~S.} \bibnamefont{Dzurak}}, \bibinfo{author}{\bibfnamefont{E.~L.} \bibnamefont{Hu}}, \bibnamefont{and} \bibinfo{author}{\bibfnamefont{J.~R.} \bibnamefont{Petta}}, \bibinfo{journal}{SCIENCE} \textbf{\bibinfo{volume}{339}}, \bibinfo{pages}{1174} (\bibinfo{year}{2013}), ISSN \bibinfo{issn}{0036-8075}.

\bibitem[{\citenamefont{Sarovar et~al.}(2010)\citenamefont{Sarovar, Ishizaki, Fleming, and Whaley}}]{WOS:000279014400020}
\bibinfo{author}{\bibfnamefont{M.}~\bibnamefont{Sarovar}}, \bibinfo{author}{\bibfnamefont{A.}~\bibnamefont{Ishizaki}}, \bibinfo{author}{\bibfnamefont{G.~R.} \bibnamefont{Fleming}}, \bibnamefont{and} \bibinfo{author}{\bibfnamefont{K.~B.} \bibnamefont{Whaley}}, \bibinfo{journal}{NATURE PHYSICS} \textbf{\bibinfo{volume}{6}}, \bibinfo{pages}{462} (\bibinfo{year}{2010}), ISSN \bibinfo{issn}{1745-2473}.

\bibitem[{\citenamefont{Amico et~al.}(2008)\citenamefont{Amico, Fazio, Osterloh, and Vedral}}]{WOS:000256528500006}
\bibinfo{author}{\bibfnamefont{L.}~\bibnamefont{Amico}}, \bibinfo{author}{\bibfnamefont{R.}~\bibnamefont{Fazio}}, \bibinfo{author}{\bibfnamefont{A.}~\bibnamefont{Osterloh}}, \bibnamefont{and} \bibinfo{author}{\bibfnamefont{V.}~\bibnamefont{Vedral}}, \bibinfo{journal}{REVIEWS OF MODERN PHYSICS} \textbf{\bibinfo{volume}{80}}, \bibinfo{pages}{517} (\bibinfo{year}{2008}), ISSN \bibinfo{issn}{0034-6861}.

\bibitem[{\citenamefont{Znidaric et~al.}(2008)\citenamefont{Znidaric, Prosen, and Prelovsek}}]{WOS:000253764100063}
\bibinfo{author}{\bibfnamefont{M.}~\bibnamefont{Znidaric}}, \bibinfo{author}{\bibfnamefont{T.}~\bibnamefont{Prosen}}, \bibnamefont{and} \bibinfo{author}{\bibfnamefont{P.}~\bibnamefont{Prelovsek}}, \bibinfo{journal}{PHYSICAL REVIEW B} \textbf{\bibinfo{volume}{77}} (\bibinfo{year}{2008}), ISSN \bibinfo{issn}{2469-9950}.

\bibitem[{\citenamefont{VonNeumann}(1932)}]{VonNeumann1932}
\bibinfo{author}{\bibfnamefont{J.}~\bibnamefont{VonNeumann}}, \emph{\bibinfo{title}{Mathematische Grundlagen der Quantenmechanik}} (\bibinfo{publisher}{Springer}, \bibinfo{year}{1932}).

\bibitem[{\citenamefont{Gao and Hatano}(2024)}]{WOS:001230903700002}
\bibinfo{author}{\bibfnamefont{J.}~\bibnamefont{Gao}} \bibnamefont{and} \bibinfo{author}{\bibfnamefont{N.}~\bibnamefont{Hatano}}, \bibinfo{journal}{PHYSICAL REVIEW RESEARCH} \textbf{\bibinfo{volume}{6}} (\bibinfo{year}{2024}).

\bibitem[{\citenamefont{Huang et~al.}(2024)\citenamefont{Huang, Xia, and Man}}]{WOS:001195261700001}
\bibinfo{author}{\bibfnamefont{R.}~\bibnamefont{Huang}}, \bibinfo{author}{\bibfnamefont{Y.-J.} \bibnamefont{Xia}}, \bibnamefont{and} \bibinfo{author}{\bibfnamefont{Z.-X.} \bibnamefont{Man}}, \bibinfo{journal}{NEW JOURNAL OF PHYSICS} \textbf{\bibinfo{volume}{26}} (\bibinfo{year}{2024}), ISSN \bibinfo{issn}{1367-2630}.

\bibitem[{\citenamefont{Ono et~al.}(2020)\citenamefont{Ono, Shevchenko, Mori, Moriyama, and Nori}}]{WOS:000577236900006}
\bibinfo{author}{\bibfnamefont{K.}~\bibnamefont{Ono}}, \bibinfo{author}{\bibfnamefont{S.~N.} \bibnamefont{Shevchenko}}, \bibinfo{author}{\bibfnamefont{T.}~\bibnamefont{Mori}}, \bibinfo{author}{\bibfnamefont{S.}~\bibnamefont{Moriyama}}, \bibnamefont{and} \bibinfo{author}{\bibfnamefont{F.}~\bibnamefont{Nori}}, \bibinfo{journal}{PHYSICAL REVIEW LETTERS} \textbf{\bibinfo{volume}{125}} (\bibinfo{year}{2020}), ISSN \bibinfo{issn}{0031-9007}.

\bibitem[{\citenamefont{Xu et~al.}(2024)\citenamefont{Xu, Jin, Neto, and de~Almeida}}]{xu2024universal}
\bibinfo{author}{\bibfnamefont{H.-G.} \bibnamefont{Xu}}, \bibinfo{author}{\bibfnamefont{J.}~\bibnamefont{Jin}}, \bibinfo{author}{\bibfnamefont{G.}~\bibnamefont{Neto}}, \bibnamefont{and} \bibinfo{author}{\bibfnamefont{N.~G.} \bibnamefont{de~Almeida}}, \bibinfo{journal}{Physical Review E} \textbf{\bibinfo{volume}{109}}, \bibinfo{pages}{014122} (\bibinfo{year}{2024}).

\bibitem[{\citenamefont{Matos et~al.}(2023)\citenamefont{Matos, de~Assis, and de~Almeida}}]{matos2023quantum}
\bibinfo{author}{\bibfnamefont{R.~Q.} \bibnamefont{Matos}}, \bibinfo{author}{\bibfnamefont{R.~J.} \bibnamefont{de~Assis}}, \bibnamefont{and} \bibinfo{author}{\bibfnamefont{N.~G.} \bibnamefont{de~Almeida}}, \bibinfo{journal}{Physical Review E} \textbf{\bibinfo{volume}{108}}, \bibinfo{pages}{054131} (\bibinfo{year}{2023}).

\bibitem[{\citenamefont{Castorene et~al.}(2024)\citenamefont{Castorene, Pe\~na, Norambuena, Ulloa, Araya, and Vargas}}]{Castorene}
\bibinfo{author}{\bibfnamefont{B.}~\bibnamefont{Castorene}}, \bibinfo{author}{\bibfnamefont{F.~J.} \bibnamefont{Pe\~na}}, \bibinfo{author}{\bibfnamefont{A.}~\bibnamefont{Norambuena}}, \bibinfo{author}{\bibfnamefont{S.~E.} \bibnamefont{Ulloa}}, \bibinfo{author}{\bibfnamefont{C.}~\bibnamefont{Araya}}, \bibnamefont{and} \bibinfo{author}{\bibfnamefont{P.}~\bibnamefont{Vargas}}, \bibinfo{journal}{Phys. Rev. E} \textbf{\bibinfo{volume}{110}}, \bibinfo{pages}{044135} (\bibinfo{year}{2024}), \urlprefix\url{https://link.aps.org/doi/10.1103/PhysRevE.110.044135}.

\bibitem[{\citenamefont{Araya et~al.}(2023)\citenamefont{Araya, Peña, Norambuena, Castorene, and Vargas}}]{technologies11060169}
\bibinfo{author}{\bibfnamefont{C.}~\bibnamefont{Araya}}, \bibinfo{author}{\bibfnamefont{F.~J.} \bibnamefont{Peña}}, \bibinfo{author}{\bibfnamefont{A.}~\bibnamefont{Norambuena}}, \bibinfo{author}{\bibfnamefont{B.}~\bibnamefont{Castorene}}, \bibnamefont{and} \bibinfo{author}{\bibfnamefont{P.}~\bibnamefont{Vargas}}, \bibinfo{journal}{Technologies} \textbf{\bibinfo{volume}{11}} (\bibinfo{year}{2023}), ISSN \bibinfo{issn}{2227-7080}, \urlprefix\url{https://www.mdpi.com/2227-7080/11/6/169}.

\bibitem[{\citenamefont{SREDNICKI}(1993)}]{WOS:A1993LQ10700003}
\bibinfo{author}{\bibfnamefont{M.}~\bibnamefont{SREDNICKI}}, \bibinfo{journal}{PHYSICAL REVIEW LETTERS} \textbf{\bibinfo{volume}{71}}, \bibinfo{pages}{666} (\bibinfo{year}{1993}), ISSN \bibinfo{issn}{0031-9007}.

\bibitem[{\citenamefont{Bhattacharya et~al.}(2013)\citenamefont{Bhattacharya, Nozaki, Takayanagi, and Ugajin}}]{WOS:000317186000001}
\bibinfo{author}{\bibfnamefont{J.}~\bibnamefont{Bhattacharya}}, \bibinfo{author}{\bibfnamefont{M.}~\bibnamefont{Nozaki}}, \bibinfo{author}{\bibfnamefont{T.}~\bibnamefont{Takayanagi}}, \bibnamefont{and} \bibinfo{author}{\bibfnamefont{T.}~\bibnamefont{Ugajin}}, \bibinfo{journal}{PHYSICAL REVIEW LETTERS} \textbf{\bibinfo{volume}{110}} (\bibinfo{year}{2013}), ISSN \bibinfo{issn}{0031-9007}.

\bibitem[{\citenamefont{Zhou et~al.}(2017)\citenamefont{Zhou, Kanoda, and Ng}}]{zhou2017quantum}
\bibinfo{author}{\bibfnamefont{Y.}~\bibnamefont{Zhou}}, \bibinfo{author}{\bibfnamefont{K.}~\bibnamefont{Kanoda}}, \bibnamefont{and} \bibinfo{author}{\bibfnamefont{T.-K.} \bibnamefont{Ng}}, \bibinfo{journal}{Reviews of Modern Physics} \textbf{\bibinfo{volume}{89}}, \bibinfo{pages}{025003} (\bibinfo{year}{2017}).

\bibitem[{\citenamefont{Mikitik and Sharlai}(2019)}]{mikitik2019magnetic}
\bibinfo{author}{\bibfnamefont{G.}~\bibnamefont{Mikitik}} \bibnamefont{and} \bibinfo{author}{\bibfnamefont{Y.~V.} \bibnamefont{Sharlai}}, \bibinfo{journal}{Journal of Low Temperature Physics} \textbf{\bibinfo{volume}{197}}, \bibinfo{pages}{272} (\bibinfo{year}{2019}).

\bibitem[{\citenamefont{Hurst and Flebus}(2022)}]{hurst2022non}
\bibinfo{author}{\bibfnamefont{H.~M.} \bibnamefont{Hurst}} \bibnamefont{and} \bibinfo{author}{\bibfnamefont{B.}~\bibnamefont{Flebus}}, \bibinfo{journal}{Journal of Applied Physics} \textbf{\bibinfo{volume}{132}} (\bibinfo{year}{2022}).

\bibitem[{\citenamefont{Bergholtz et~al.}(2021)\citenamefont{Bergholtz, Budich, and Kunst}}]{bergholtz2021exceptional}
\bibinfo{author}{\bibfnamefont{E.~J.} \bibnamefont{Bergholtz}}, \bibinfo{author}{\bibfnamefont{J.~C.} \bibnamefont{Budich}}, \bibnamefont{and} \bibinfo{author}{\bibfnamefont{F.~K.} \bibnamefont{Kunst}}, \bibinfo{journal}{Reviews of Modern Physics} \textbf{\bibinfo{volume}{93}}, \bibinfo{pages}{015005} (\bibinfo{year}{2021}).

\bibitem[{\citenamefont{Mishra and Singh}(2024)}]{Bell1}
\bibinfo{author}{\bibfnamefont{S.}~\bibnamefont{Mishra}} \bibnamefont{and} \bibinfo{author}{\bibfnamefont{R.}~\bibnamefont{Singh}}, \bibinfo{journal}{Physics Open} \textbf{\bibinfo{volume}{18}}, \bibinfo{pages}{100199} (\bibinfo{year}{2024}), ISSN \bibinfo{issn}{2666-0326}, \urlprefix\url{https://www.sciencedirect.com/science/article/pii/S2666032623000649}.

\bibitem[{\citenamefont{Barbieri et~al.}(2004)\citenamefont{Barbieri, De~Martini, Di~Nepi, and Mataloni}}]{Werner1}
\bibinfo{author}{\bibfnamefont{M.}~\bibnamefont{Barbieri}}, \bibinfo{author}{\bibfnamefont{F.}~\bibnamefont{De~Martini}}, \bibinfo{author}{\bibfnamefont{G.}~\bibnamefont{Di~Nepi}}, \bibnamefont{and} \bibinfo{author}{\bibfnamefont{P.}~\bibnamefont{Mataloni}}, \bibinfo{journal}{Phys. Rev. Lett.} \textbf{\bibinfo{volume}{92}}, \bibinfo{pages}{177901} (\bibinfo{year}{2004}), \urlprefix\url{https://link.aps.org/doi/10.1103/PhysRevLett.92.177901}.

\bibitem[{\citenamefont{V\'ertesi}(2008)}]{Werner2}
\bibinfo{author}{\bibfnamefont{T.}~\bibnamefont{V\'ertesi}}, \bibinfo{journal}{Phys. Rev. A} \textbf{\bibinfo{volume}{78}}, \bibinfo{pages}{032112} (\bibinfo{year}{2008}), \urlprefix\url{https://link.aps.org/doi/10.1103/PhysRevA.78.032112}.

\bibitem[{\citenamefont{Wiesniak et~al.}(2005)\citenamefont{Wiesniak, Vedral, and Brukner}}]{WOS:000234347200003}
\bibinfo{author}{\bibfnamefont{M.}~\bibnamefont{Wiesniak}}, \bibinfo{author}{\bibfnamefont{V.}~\bibnamefont{Vedral}}, \bibnamefont{and} \bibinfo{author}{\bibfnamefont{C.}~\bibnamefont{Brukner}}, \bibinfo{journal}{NEW JOURNAL OF PHYSICS} \textbf{\bibinfo{volume}{7}} (\bibinfo{year}{2005}), ISSN \bibinfo{issn}{1367-2630}.

\bibitem[{\citenamefont{De~Chiara et~al.}(2006)\citenamefont{De~Chiara, Brukner, Fazio, Palma, and Vedral}}]{WOS:000238204500001}
\bibinfo{author}{\bibfnamefont{G.}~\bibnamefont{De~Chiara}}, \bibinfo{author}{\bibfnamefont{C.}~\bibnamefont{Brukner}}, \bibinfo{author}{\bibfnamefont{R.}~\bibnamefont{Fazio}}, \bibinfo{author}{\bibfnamefont{G.}~\bibnamefont{Palma}}, \bibnamefont{and} \bibinfo{author}{\bibfnamefont{V.}~\bibnamefont{Vedral}}, \bibinfo{journal}{NEW JOURNAL OF PHYSICS} \textbf{\bibinfo{volume}{8}} (\bibinfo{year}{2006}), ISSN \bibinfo{issn}{1367-2630}.

\bibitem[{\citenamefont{Rappoport et~al.}(2007)\citenamefont{Rappoport, Ghivelder, Fernandes, Guimaraes, and Continentino}}]{WOS:000244532600066}
\bibinfo{author}{\bibfnamefont{T.~G.} \bibnamefont{Rappoport}}, \bibinfo{author}{\bibfnamefont{L.}~\bibnamefont{Ghivelder}}, \bibinfo{author}{\bibfnamefont{J.~C.} \bibnamefont{Fernandes}}, \bibinfo{author}{\bibfnamefont{R.~B.} \bibnamefont{Guimaraes}}, \bibnamefont{and} \bibinfo{author}{\bibfnamefont{M.~A.} \bibnamefont{Continentino}}, \bibinfo{journal}{PHYSICAL REVIEW B} \textbf{\bibinfo{volume}{75}} (\bibinfo{year}{2007}), ISSN \bibinfo{issn}{2469-9950}.

\bibitem[{\citenamefont{Sachdev}(1999)}]{qpt1}
\bibinfo{author}{\bibfnamefont{S.}~\bibnamefont{Sachdev}}, \bibinfo{journal}{Physics world} \textbf{\bibinfo{volume}{12}}, \bibinfo{pages}{33} (\bibinfo{year}{1999}).

\bibitem[{\citenamefont{Fu-Wu and Xiang-Mu}(2012)}]{qpt2}
\bibinfo{author}{\bibfnamefont{M.}~\bibnamefont{Fu-Wu}} \bibnamefont{and} \bibinfo{author}{\bibfnamefont{K.}~\bibnamefont{Xiang-Mu}}, \bibinfo{journal}{COMMUNICATIONS IN THEORETICAL PHYSICS} \textbf{\bibinfo{volume}{57}}, \bibinfo{pages}{999} (\bibinfo{year}{2012}), ISSN \bibinfo{issn}{0253-6102}.

\bibitem[{\citenamefont{Werlang et~al.}(2013)\citenamefont{Werlang, Ribeiro, and Rigolin}}]{qpt3}
\bibinfo{author}{\bibfnamefont{T.}~\bibnamefont{Werlang}}, \bibinfo{author}{\bibfnamefont{G.~A.~P.} \bibnamefont{Ribeiro}}, \bibnamefont{and} \bibinfo{author}{\bibfnamefont{G.}~\bibnamefont{Rigolin}}, \bibinfo{journal}{INTERNATIONAL JOURNAL OF MODERN PHYSICS B} \textbf{\bibinfo{volume}{27}} (\bibinfo{year}{2013}), ISSN \bibinfo{issn}{0217-9792}.

\bibitem[{\citenamefont{Vojta}(2000)}]{qpt4}
\bibinfo{author}{\bibfnamefont{T.}~\bibnamefont{Vojta}}, \bibinfo{journal}{ANNALEN DER PHYSIK} \textbf{\bibinfo{volume}{9}}, \bibinfo{pages}{403} (\bibinfo{year}{2000}), ISSN \bibinfo{issn}{0003-3804}.

\bibitem[{\citenamefont{Shi-Rong et~al.}(2010)\citenamefont{Shi-Rong, Yun-Jie, and Zhong-Xiao}}]{qpt5}
\bibinfo{author}{\bibfnamefont{C.}~\bibnamefont{Shi-Rong}}, \bibinfo{author}{\bibfnamefont{X.}~\bibnamefont{Yun-Jie}}, \bibnamefont{and} \bibinfo{author}{\bibfnamefont{M.}~\bibnamefont{Zhong-Xiao}}, \bibinfo{journal}{CHINESE PHYSICS B} \textbf{\bibinfo{volume}{19}} (\bibinfo{year}{2010}), ISSN \bibinfo{issn}{1674-1056}.

\bibitem[{\citenamefont{Vojta}(2003)}]{qpt6}
\bibinfo{author}{\bibfnamefont{M.}~\bibnamefont{Vojta}}, \bibinfo{journal}{REPORTS ON PROGRESS IN PHYSICS} \textbf{\bibinfo{volume}{66}}, \bibinfo{pages}{2069} (\bibinfo{year}{2003}), ISSN \bibinfo{issn}{0034-4885}.

\bibitem[{\citenamefont{Xiao-Dong et~al.}(2013)\citenamefont{Xiao-Dong, Shou-Sheng, and Bai-Qi}}]{qpt7}
\bibinfo{author}{\bibfnamefont{T.}~\bibnamefont{Xiao-Dong}}, \bibinfo{author}{\bibfnamefont{H.}~\bibnamefont{Shou-Sheng}}, \bibnamefont{and} \bibinfo{author}{\bibfnamefont{J.}~\bibnamefont{Bai-Qi}}, \bibinfo{journal}{COMMUNICATIONS IN THEORETICAL PHYSICS} \textbf{\bibinfo{volume}{59}}, \bibinfo{pages}{146} (\bibinfo{year}{2013}), ISSN \bibinfo{issn}{0253-6102}.

\bibitem[{\citenamefont{Allahverdyan et~al.}(2006)\citenamefont{Allahverdyan, Balian, and Nieuwenhuizen}}]{qpt8}
\bibinfo{author}{\bibfnamefont{A.}~\bibnamefont{Allahverdyan}}, \bibinfo{author}{\bibfnamefont{R.}~\bibnamefont{Balian}}, \bibnamefont{and} \bibinfo{author}{\bibfnamefont{T.}~\bibnamefont{Nieuwenhuizen}}, in \emph{\bibinfo{booktitle}{QUANTUM THEORY: RECONSIDERATION OF FOUNDATIONS - 3}}, edited by \bibinfo{editor}{\bibfnamefont{G.}~\bibnamefont{Adenier}}, \bibinfo{editor}{\bibfnamefont{A.}~\bibnamefont{Khrennikov}}, \bibnamefont{and} \bibinfo{editor}{\bibfnamefont{T.}~\bibnamefont{nieuwenhuizen}} (\bibinfo{organization}{Swedish Res Council; Eu-Network Quantum Probability; Vaxjo Univ, Profile Math Modelling}, \bibinfo{year}{2006}), vol. \bibinfo{volume}{810} of \emph{\bibinfo{series}{AIP Conference Proceedings}}, pp. \bibinfo{pages}{47+}, ISBN \bibinfo{isbn}{0-7354-0301-5}, ISSN \bibinfo{issn}{0094-243X}, \bibinfo{note}{international Conference on Quantum Theory - Reconsideration of Foundations -3, Vaxjo, SWEDEN, JUN 06-11, 2005}.

\bibitem[{\citenamefont{Bahmani et~al.}(2020)\citenamefont{Bahmani, Najarbashi, and Tavana}}]{qpt9}
\bibinfo{author}{\bibfnamefont{H.}~\bibnamefont{Bahmani}}, \bibinfo{author}{\bibfnamefont{G.}~\bibnamefont{Najarbashi}}, \bibnamefont{and} \bibinfo{author}{\bibfnamefont{A.}~\bibnamefont{Tavana}}, \bibinfo{journal}{Physica Scripta} \textbf{\bibinfo{volume}{95}}, \bibinfo{pages}{055701} (\bibinfo{year}{2020}), \urlprefix\url{https://dx.doi.org/10.1088/1402-4896/ab606e}.

\bibitem[{\citenamefont{Throckmorton and Das~Sarma}(2021)}]{qpt10}
\bibinfo{author}{\bibfnamefont{R.~E.} \bibnamefont{Throckmorton}} \bibnamefont{and} \bibinfo{author}{\bibfnamefont{S.}~\bibnamefont{Das~Sarma}}, \bibinfo{journal}{Phys. Rev. B} \textbf{\bibinfo{volume}{103}}, \bibinfo{pages}{165431} (\bibinfo{year}{2021}), \urlprefix\url{https://link.aps.org/doi/10.1103/PhysRevB.103.165431}.

\end{thebibliography}

\clearpage
\onecolumngrid

\end{document}